\documentclass[essd, manuscript]{copernicus}

\usepackage{amsmath,amssymb}
\usepackage{booktabs}
\usepackage{array}
\usepackage{multirow}

\usepackage{siunitx}
\usepackage{graphicx}
\usepackage{xcolor}
\usepackage[round,authoryear]{natbib}

\begin{document}

\extrafloats{80}

\title{TailWeather: from tail to extremes, a global climatological dataset for machine-learning weather forecasting}

\Author[1,2][zhisong.liu@lut.fi]{Zhi-Song}{Liu}
\Author[1,2,3]{Michael}{Boy}
\Author[4]{Risto}{Makkonen}

\affil[1]{Department of Computational Engineering, LUT University, Finland}
\affil[2]{Atmospheric Modelling Center, Lahti (AMC-Lahti), Finland}
\affil[3]{Institute for Atmospheric and Earth System Research (INAR),
University of Helsinki, Finland}
\affil[4]{Finnish Meteorological Institute (FMI), Finland}

\runningtitle{TailWeather: climatological tail events for weather forecasting}
\runningauthor{Liu et al.}

\received{}
\pubdiscuss{}
\revised{}
\accepted{}
\published{}

\firstpage{1}
\maketitle


\begin{abstract}
Weather forecasting models commonly use the average forecast skill for short-range forecast. However, it does not necessarily imply skill in the tails of the weather distribution. Evaluating tail events requires a consistently defined target with broad spatial and temporal coverage. Disaster catalogs record societal consequences, but are sparse and reporting-dependent; climatological tails describe unusual weather without necessarily implying harm. We present TailWeather, a global, \ang{0.25}, land-only dataset derived from ERA5, covering 1981--2022 and extending into January 2023. It labels heatwaves, cold waves, heavy precipitation, and extreme wind daily, and meteorological drought monthly. Each event has an ordinal severity tier and a numerical intensity score referenced to the local 1991--2020 climate. The scores support alternative thresholds within their stored resolution and valid domain. Comparison with documented disasters shows greater impact enrichment towards stricter tails, with differences among hazards and substantial gaps in catalog coverage. Forecast examples illustrate how low average errors can coexist with weak event detection, particularly for wind. TailWeather provides a reusable physical target for studying and evaluating extremes, while complementing the information in disaster catalogs.

\end{abstract}

\introduction
\label{sec:intro}

\begin{figure*}[tbp]
\centering
\includegraphics[width=\textwidth]{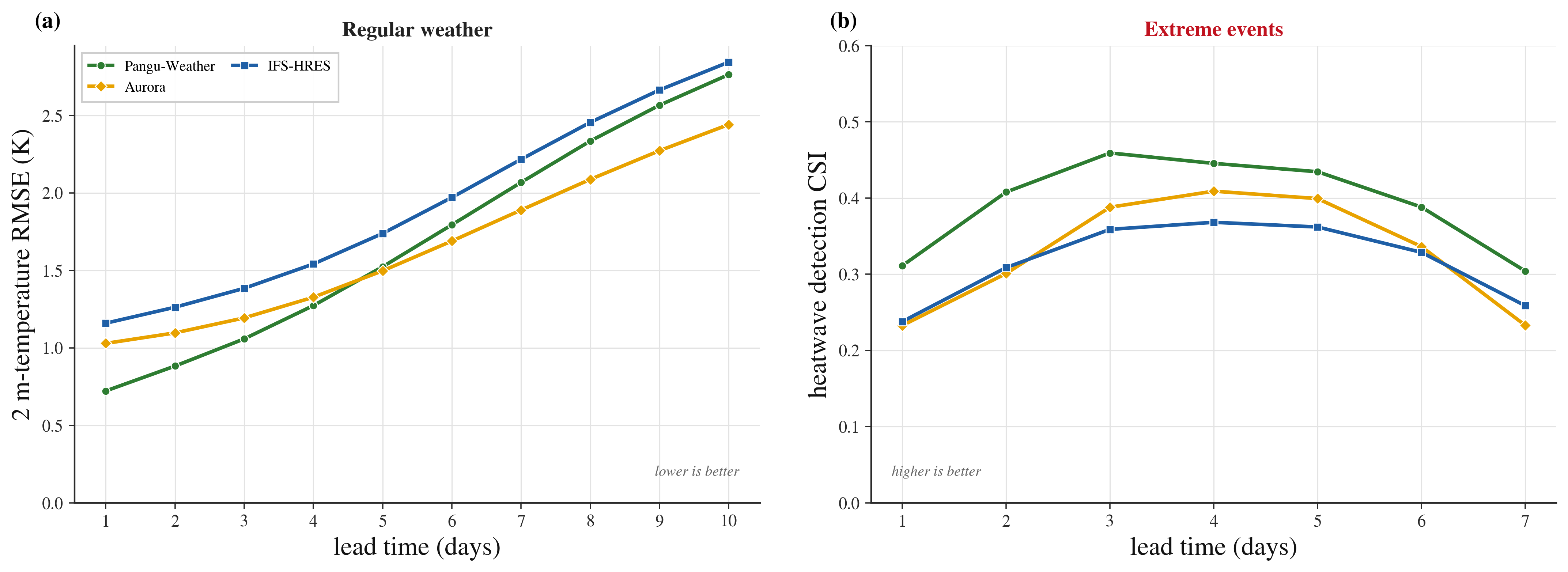}
\caption{\textbf{Strong average skill does not imply strong extreme skill.}
Three leading forecast systems evaluated on WeatherBench2 for 2022. (a) Regular-weather skill: 2 m-temperature root-mean-square error (RMSE) against ERA5 grows smoothly with lead time, from about 1 K at day~1 to 3 K at day 10, close to the state of the art. (b) Extreme-event skill: heatwave-detection critical success index (CSI) against the tail label, for the same systems, peaks below 0.5 near day 3 and then declines. Each plot uses the standard metric of its regime (RMSE for the field regression, CSI for event detection).}
\label{fig:reg-vs-ext}
\end{figure*}

AI weather models now match or exceed operational numerical weather prediction on
average scores at a fraction of the computational cost
\citep{bi2023pangu, lam2023graphcast, price2025gencast, rasp2024weatherbench2}.
But average scores are dominated by ordinary weather. They do not answer the
question that matters for extremes: does a forecast identify the rare conditions
at a particular place and time? Figure~\ref{fig:reg-vs-ext} illustrates the
gap. Models with low temperature error can still have limited skill at detecting
heatwave tail events. Answering this question requires a target that is defined
at every land cell and time step.

Two kinds of data describe extremes, and neither can replace the other. EM-DAT records reported disasters \citep{emdat}, while the NOAA Storm Events Database combines descriptions of weather events and their impacts \citep{noaa_stormevents}. Such records reflect exposure, vulnerability, and reporting practice as well as the weather itself. They are sparse and unevenly distributed, so they cannot evaluate a forecast everywhere. Other archives serve different purposes: IBTrACS, for example, compiles tropical-cyclone tracks and characteristics \citep{knapp2010ibtracs}. Related benchmarks serve complementary purposes: HR-Extreme provides high-resolution regional extreme-weather cases \citep{ran2025hrextreme}; Extreme Weather Bench combines curated high-impact cases with observations and evaluation tools \citep{mcgovern2026ewb}; and ExEBench supports multiple event categories and tasks using several data sources \citep{zhao2025exebench}.

Climatological tails answer a different question. A value is extreme when it is
rare relative to the local climate. This approach underlies the Expert Team on
Climate Change Detection and Indices, percentile-and-persistence heatwave
definitions, the Standardized Precipitation Index, and extreme-value methods
\citep{zhang2011indices, perkins2013heatwave, mckee1993spi, coles2001evt}. The definition can be applied consistently, but rarity alone does not imply harm. A
warm day in an unpopulated region can be statistically extreme without causing a
disaster; a damaging tornado may be too small for a \ang{0.25} reanalysis to
resolve. The useful approach is therefore to keep physical tails and recorded
impacts separate, then measure their relationship.

We present TailWeather, a global, land-only dataset of climatological tail
events covering 1981--2022 and part of January 2023. It provides daily labels for heatwave, cold wave, heavy precipitation, and extreme wind, and monthly labels for meteorological drought. Each label has a severity tier and an intensity score. Within the stored score resolution and valid domain, users can choose thresholds and derive compound tails without rebuilding the reference climatology. TailWeather is designed as a dense physical target for weather models, while impact records provide an independent check of its societal
relevance.

To summarize the contributions: First, TailWeather gives a consistent global
target for five hazards. Second, higher tail intensity is more often associated
with recorded impacts, but the useful threshold differs by hazard and impact
records cannot assess much of the world. Third, compound tails do not
consistently improve that association. Finally, the dataset reveals weaknesses
in AI extreme-event prediction that average forecast scores conceal, and it can
also be used as a training signal. The sharpening-head experiment is included as
an example application, not as a general training method. TailWeather complements these resources with a long global record and consistently applied definitions, while it does not resolve convective hazards or measure impacts directly. The following sections describe the labels, test their relationship with documented disasters, and show two example uses in AI weather forecasting.

\begin{figure*}[tbp]
\centering
\includegraphics[width=0.95\textwidth]{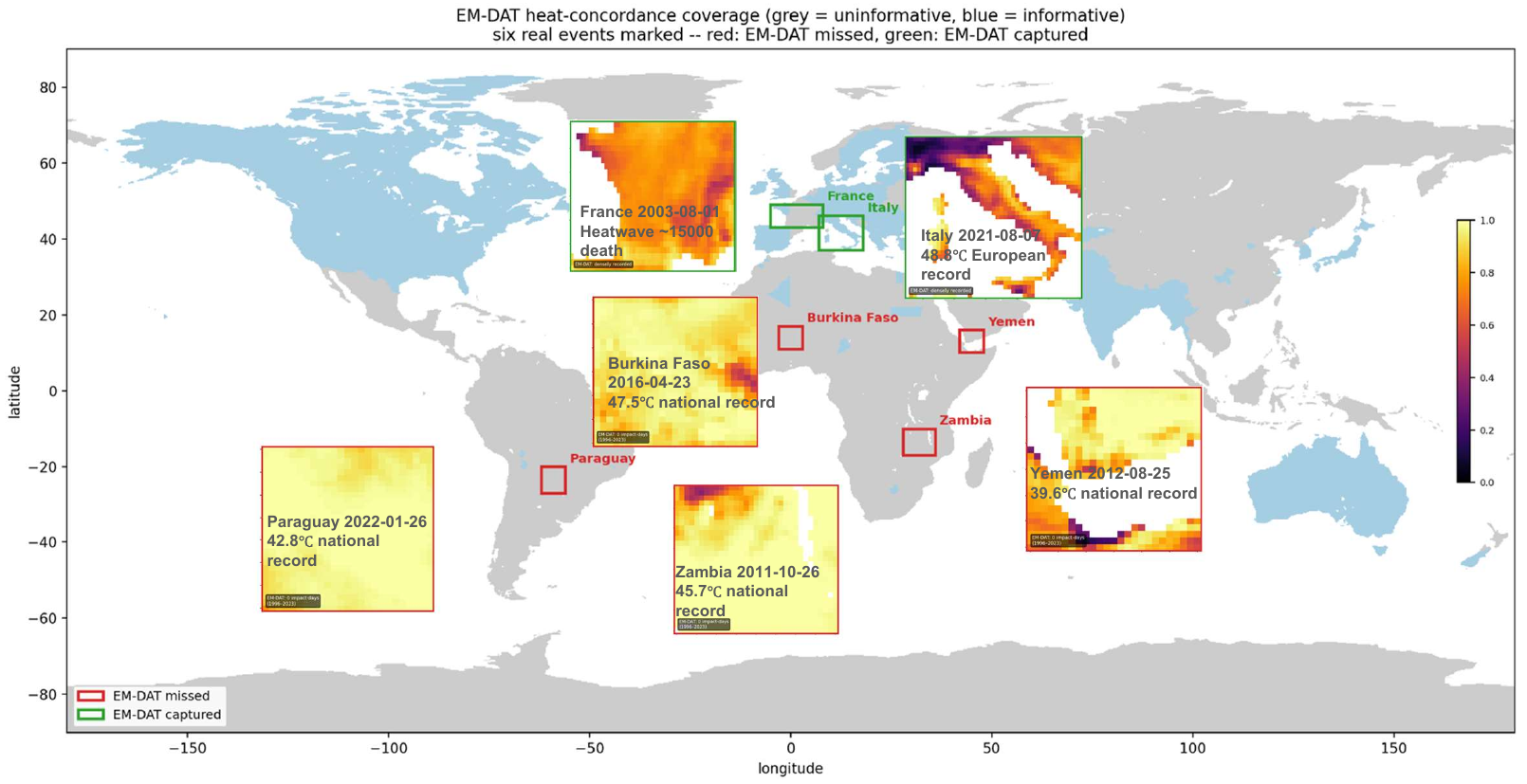}
\caption{Heat-impact catalog coverage and selected heat events. Grey cells have fewer than five recorded impact days in the nominal 1996--2023 comparison period; blue cells meet this descriptive coverage screen. Red boxes locate four station or national heat records (Paraguay, Zambia, Burkina Faso, and Yemen) without a matched EM-DAT entry in the comparison; green boxes locate the 2003 France and 2021 Italy heatwaves. These examples illustrate differences in catalog coverage and do not establish the completeness of either source.}
\label{fig:teaser-coverage}
\end{figure*}

\section{TailWeather: a dense tail-event benchmark}
\label{sec:source}

All labels derive from ERA5 reanalysis \citep{hersbach2020era5}, accessed
through the WeatherBench2 analysis-ready zarr store
\citep{rasp2024weatherbench2}, specifically
\texttt{1959-2023\_01\_10-wb13-6h-1440x721\_with\_derived\_variables.zarr}.
Using one source keeps input provenance consistent across hazards, while retaining ERA5's variable-specific biases.

\noindent\textbf{Grid and period.}
The labels use a regular \ang{0.25} grid ($721\times1440$). The reference climate is 1991--2020. The record extends from 1981 into January 2023; the terminal year is incomplete. Hazard-specific end dates and processing conventions are given in Appendix~\ref{app:processing}.

\noindent\textbf{Land mask.}
The ERA5 land--sea mask at $\ge0.5$ selects $351\,848$ land cells, including ice sheets. This is 33.9\,\% of grid points, not surface area, because latitude--longitude cells have unequal areas. Ocean labels are zero and continuous fields are missing; users must retain the land and validity masks.

\noindent\textbf{Severity and intensity.}
The severity tier $s\in\{0,1,2,3\}$ denotes none, moderate, severe, or extreme conditions within each hazard. The intensity score $I\in[0,1]$ increases towards the relevant tail. Daily hazards use discretized percentile-rank scores; drought uses a scaled SPI score that is missing outside drought events. A common score threshold therefore need not represent the same rarity across hazards. The score definitions, precision, and limits on re-thresholding are described in Appendices~\ref{app:defs} and \ref{app:processing}.

\noindent\textbf{Empirical versus parametric tails.}
Temperature, precipitation, and wind thresholds are estimated empirically. Drought uses a gamma-based SPI transformation for three-month precipitation totals, with a probability mass at zero. Both approaches retain uncertainty from the finite reference record.

\noindent\textbf{The five hazards in brief.}
Each hazard applies a percentile-and-persistence rule to one ERA5 variable
against its local 1991--2020 climatology; Table~\ref{tab:overview} lists the
variable, threshold, and persistence for each, and Appendix~\ref{app:defs} gives
the complete definitions. In outline: a heatwave (cold wave) is a run of at least
three consecutive days above the local 90th (below the 10th) percentile of daily
maximum (minimum) \SI{2}{\meter} temperature \citep{perkins2013heatwave}; heavy
precipitation is a wet day exceeding the local 95th wet-day percentile, with a separate long-wet-spell diagnostic \citep{zolina2010precip}; extreme wind is a day exceeding the local
98th percentile of daily-maximum \SI{10}{\meter} wind speed, with
Beaufort-anchored severity tiers; and meteorological drought is a 3-month
Standardized Precipitation Index at or below $-1$, the dataset's only parametric
(gamma-fitted) tail \citep{mckee1993spi}.

\noindent\textbf{Prevalence and coverage.}
Table~\ref{tab:overview} summarises the five hazards and their prevalence (the
event-day rate, or fraction of land cell-days flagged $s\ge 1$), which depends on the thresholds, persistence, and temporal dependence of the weather. Only complete periods should be used for annual comparisons. For AI model development, we provide the dataset split scheme: a temporal split of 1981--2020 (train), 2021 (validation), and 2022 and available January 2023 dates (test) keeps the 1991--2020 climatological reference inside the training span, and the training years contain $\sim\!5.13\times10^{9}$ labelled land cell-days. Annual frequency, severity-tier distributions, a comparative per-hazard profile, and the frequency evolution under the fixed reference are given in Appendix~\ref{app:characteristics}.

\begin{table*}[tbp]
\centering
\caption{Overview of the five labelled hazards. All labels derive from ERA5 at
\ang{0.25} over land, reference period 1991--2020, with the ordinal tier
convention $\{0\,\text{none},1\,\text{moderate},2\,\text{severe},
3\,\text{extreme}\}$. $^\dagger$Heavy precipitation has a separate long-wet-spell diagnostic; its inspected severity mask follows the heavy-day flag. $^\ddagger$Drought is computed
monthly and optionally broadcast to daily. $^\S$Empty by construction: the
R95p criterion places every flagged day at or above the 95th wet-day
percentile, so the moderate bin $[0.90,0.95)$ cannot be populated
(Sect.~\ref{sec:precip}).}
\label{tab:overview}
\small
\setlength{\tabcolsep}{5pt}
\renewcommand{\arraystretch}{1.25}
\resizebox{\linewidth}{!}{
\begin{tabular}{@{}l l l l l l@{}}
\toprule
\textbf{Property} & \textbf{Heatwave} & \textbf{Cold wave} &
\textbf{Heavy precip.} & \textbf{Extreme wind} & \textbf{Drought} \\
\midrule
Source variable & $\mathrm{TX}$ (\SI{2}{\meter} max) & $\mathrm{TN}$ (\SI{2}{\meter} min) &
total precip. & \SI{10}{\meter} wind & precip. accum. \\
Threshold method & 90th-percentile climatology & 10th-percentile climatology &
95th-percentile wet day; spell diagnostic & 98th-percentile climatology & SPI-3 gamma \\
Tail type & empirical & empirical & empirical & empirical & parametric \\
Min.\ duration & 3 days & 3 days & 1 day$^\dagger$ & 1 day & monthly \\
Native temporal res. & daily & daily & daily & daily & monthly$^\ddagger$ \\
Severity basis & score bins & score bins & score bins & Beaufort bands & SPI bands \\
Reference & \citet{perkins2013heatwave} & \citet{perkins2013heatwave} &
\citet{zolina2010precip} & WMO Beaufort & \citet{mckee1993spi} \\
\midrule
\multicolumn{6}{@{}l}{\textit{Prevalence (land cell-days, 1991--2020)}}\\
Event-day rate (\%) & 4.566 & 4.008 & 1.209 & 1.996 & 15.934 \\
\% moderate ($s{=}1$) & 1.815 & 1.643 & 0.000$^\S$ & 1.931 & 9.359 \\
\% severe ($s{=}2$) & 2.104 & 1.833 & 0.964 & 0.062 & 4.448 \\
\% extreme ($s{=}3$) & 0.648 & 0.532 & 0.246 & 0.003 & 2.127 \\
\bottomrule
\end{tabular}}
\end{table*}

\begin{figure*}[tbp]
\centering
\includegraphics[width=\textwidth]{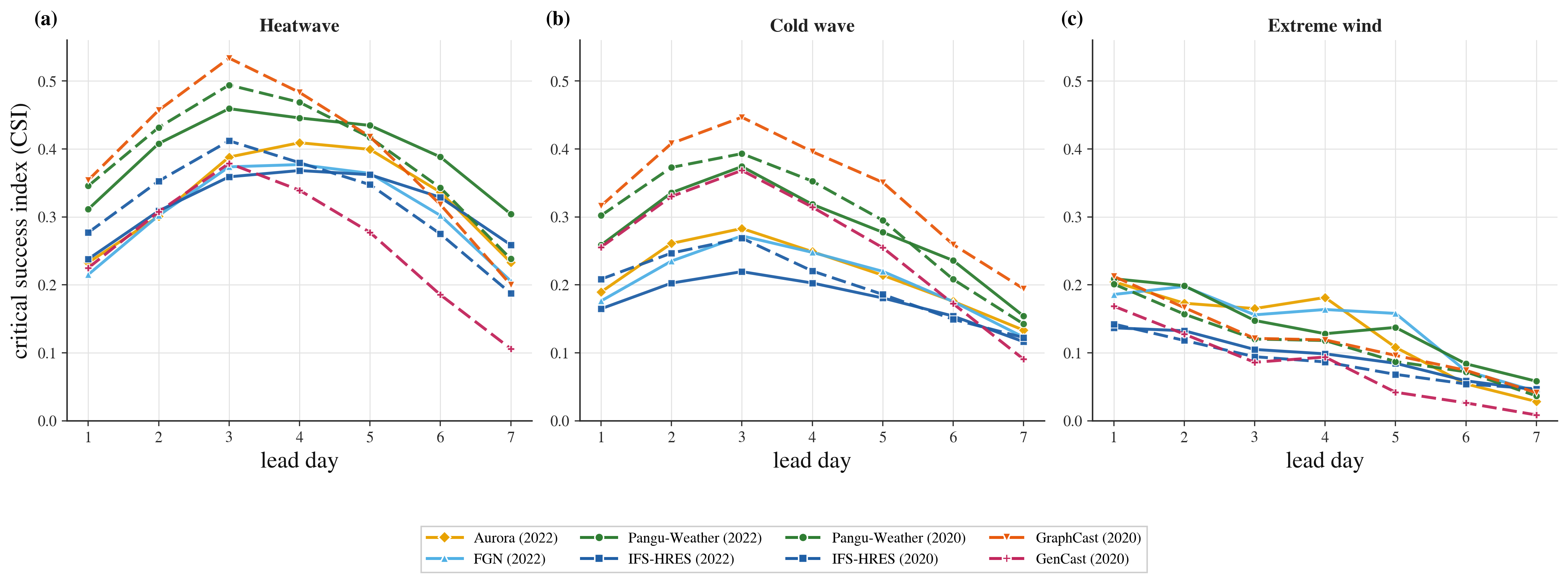}
\caption{\textbf{Extreme-event detection skill versus lead time on AI methods.} Critical success index (CSI, the higher the better) of six forecast systems, including IFS-HRES, scored against the tail labels, for the three day-resolved hazards. Solid lines are the 2022 methods (Aurora, FGN, Pangu-Weather, IFS-HRES), dashed the 2020 methods (GraphCast, GenCast, Pangu-Weather, IFS-HRES); Pangu-Weather and IFS-HRES appear in both as cross-year anchors. CSI peaks near day 3 for temperature hazards and is lowest and decays fastest for extreme wind.}
\label{fig:wb2}
\end{figure*}

\section{Example application: TailWeather reveals an extreme-skill gap}
\label{sec:aibench}

We first use TailWeather as an evaluation target. Strong average forecast skill
does not guarantee that a model predicts rare conditions well. In
Fig.~\ref{fig:reg-vs-ext}, three leading systems predict 2~m temperature to
within about \SI{1}{\kelvin} at day~1, yet their heatwave detection skill,
measured by the critical success index (CSI), remains below 0.5 even at its
day-3 peak. Average grid-point errors are dominated by ordinary weather and
cannot reveal this difference.

We test six published systems, Pangu-Weather, GraphCast, GenCast, Aurora,
FGN, and the high-resolution Integrated Forecasting System
(IFS-HRES), against the day-resolved temperature and wind labels
\citep{bi2023pangu,lam2023graphcast,price2025gencast,aurora,fgn,ecmwf_ifs_hres}.
The available forecast archives cover two test years, with Pangu-Weather and
IFS-HRES providing anchors in both. Heavy precipitation is not included because
these archives do not provide a comparable precipitation forecast, and drought
is monthly. In this comparison (Fig.~\ref{fig:wb2}), tail-event skill differs strongly by hazard. It is highest for temperature and lowest for extreme wind for every model and lead time. The diagnostic plot gives complementary information: models tend to miss temperature tails but over-predict wind tails.
A further comparison with recorded wind-disaster footprints shows that
statistical-tail detection and disaster capture are different evaluation
questions. TailWeather therefore adds a target that average forecast scores and
sparse disaster records cannot provide on their own.

\begin{figure*}[tbp]
\centering
\begin{minipage}[t]{0.49\textwidth}
\centering
\includegraphics[width=\linewidth]{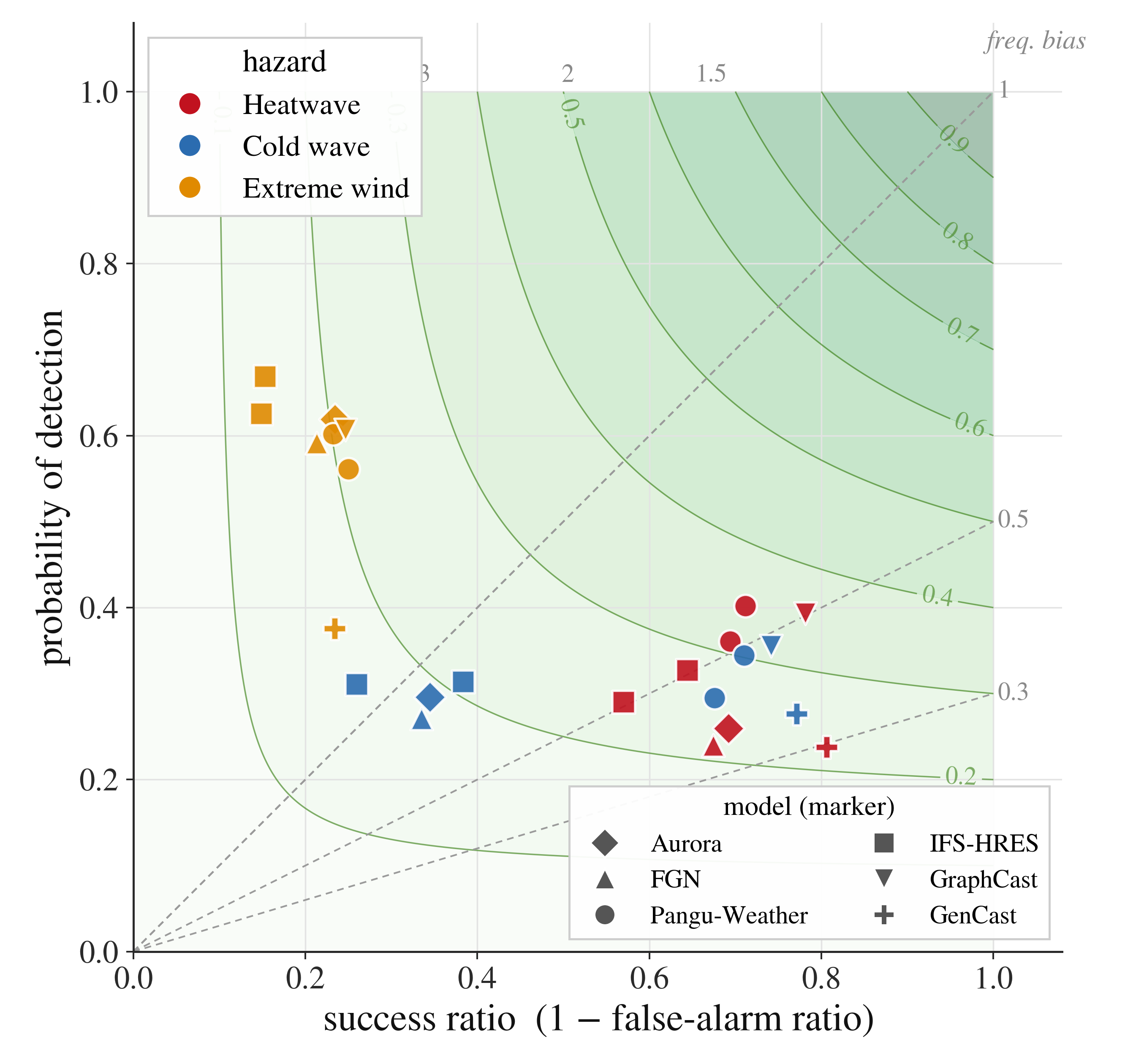}
\end{minipage}\hfill
\begin{minipage}[t]{0.49\textwidth}
\centering
\includegraphics[width=\linewidth]{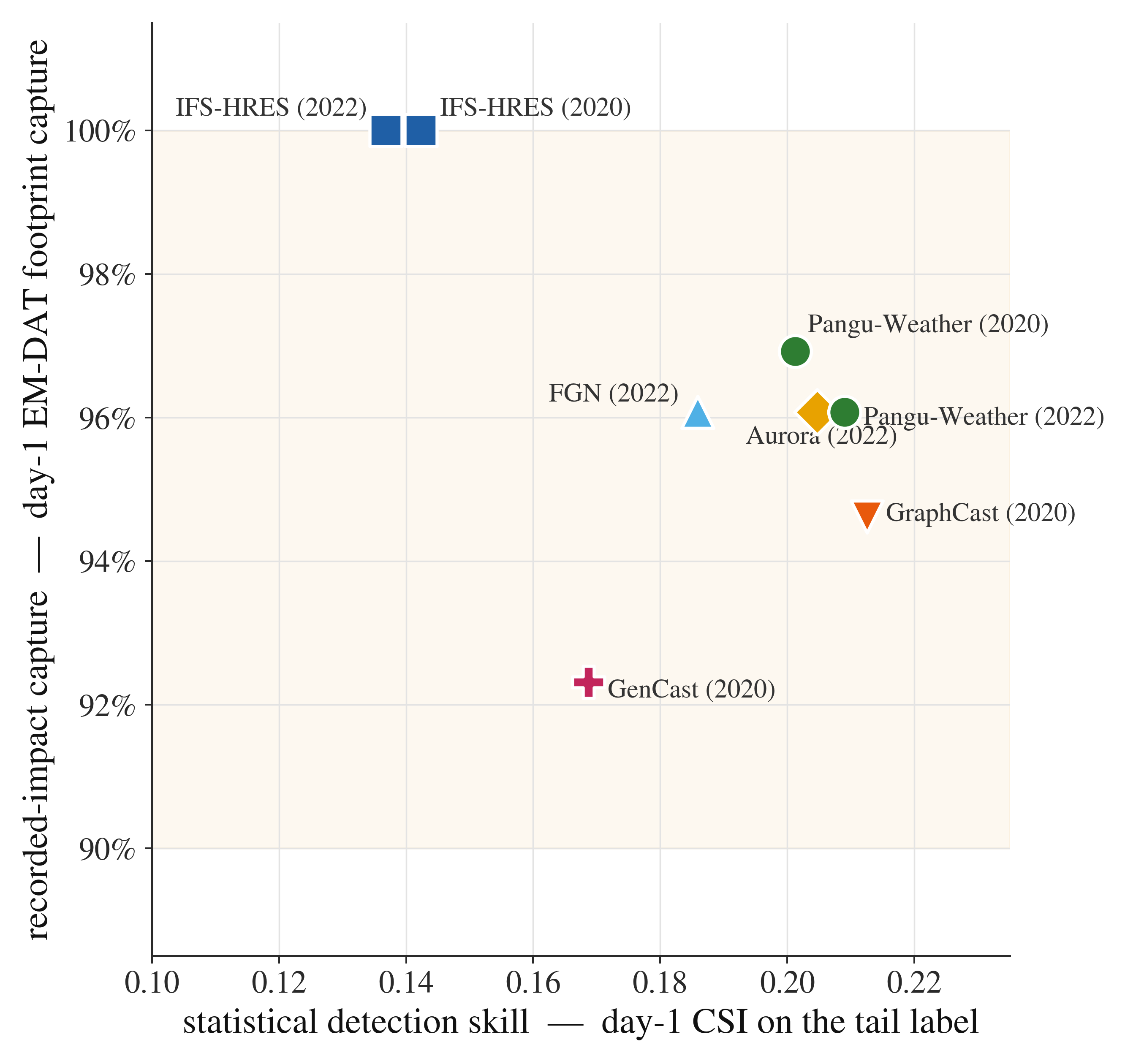}
\end{minipage}
\caption{\textbf{Two views of forecast performance.} (a) Day-1 detection
performance against the tail labels, shown as probability of detection and
success ratio (green CSI isolines; grey frequency-bias rays). Models tend to
under-predict temperature tails and over-predict extreme-wind tails. (b)
Tail-label detection skill and recorded-impact capture do not rank the six
systems in the same order for extreme wind, because they measure different
targets.}
\label{fig:wb2-diagnostics}
\end{figure*}

\section{Relationship between tail events and documented disasters}
\label{sec:quality}

TailWeather labels describe physically unusual weather, not the damage that it
causes. Their definitions are applied consistently, but they need not match documented disasters. Here we measure that relationship as a plausibility check, rather than a validation of ERA5 meteorological accuracy.
We first show where the impact record is sufficient for the comparison, then
ask whether the labels overlap recorded disasters and how that overlap changes
as the tail threshold becomes stricter.

\subsection{Design}
\label{sec:concordance-design}
We compare the labels with EM-DAT, an independently compiled global disaster
catalog \citep{emdat}, for 1996--2023. We place each record on the same
\ang{0.25} ERA5 grid as TailWeather, using sub-national boundaries from the
Geocoded Disasters dataset (GDIS) where available and country boundaries
otherwise \citep{rosvold2021gdis}. A multi-day record covers its full reported period, with a $\pm2$-day allowance for differences between local reporting dates and UTC days. We match heat, cold, wind, and drought records to their like-named labels. Flood records are matched to heavy precipitation, but this is less direct: flooding also depends on upstream rainfall, wet soils, and river routing. Table~\ref{tab:catalog} gives the available record counts and geographic detail. Matching rules and uncertainty limitations are detailed in Appendix~\ref{app:concordance-detail}.

\begin{table}[htbp]
\centering
\caption{Impact-catalog coverage, 1996--2023. ``Sub-national'' is the
percentage of records carrying administrative-unit rather than country-level
geometry. Heatwave's low share reflects that much of its record consists of
the 2022 European heatwave, entered as one row per affected country and
falling outside the sub-national geometry source's 2018 coverage limit.}
\label{tab:catalog}
\small
\setlength{\tabcolsep}{4pt}
\renewcommand{\arraystretch}{1.2}
\begin{tabular}{@{}l S S S@{}}
\toprule
\textbf{Hazard} & {Events} & {Sub-national (\%)} & {Deaths} \\
\midrule
Heatwave      & 219  & 52 & 258696 \\
Cold wave     & 288  & 93 & 15076 \\
Extreme wind  & 2495 & 73 & 220442 \\
Heavy precip. & 3966 & 73 & 130382 \\
Drought       & 402  & 75 & 24170 \\
\bottomrule
\end{tabular}
\end{table}

For each threshold $q$, we report two complementary measures. Catalog precision asks how often a tail cell-day overlaps a disaster record; catalog recall asks how often a recorded-impact cell-day is a tail. An unmatched cell-day does not necessarily lack an impact. We use a 15-day block bootstrap to account for short-term temporal dependence; its limitations are discussed in Appendix~\ref{app:concordance-detail}. The full definition is in
Appendix~\ref{app:concordance-detail}.

\subsection{Documented disaster records cover only part of the world}
\label{sec:coverage}

Before asking how closely the tail labels and EM-DAT records match, we ask
where the comparison is possible. For each hazard, we exclude land cells with
fewer than five recorded impact days over 1996--2023; this provides a simple coverage screen, although five days may belong to a single disaster. We use the same rule for the
per-cell maps in Sect.~\ref{sec:spatial}, then ask what fraction of land area
and population falls inside the excluded area. Figure~\ref{fig:teaser-coverage}
illustrates the coverage gaps and selected event locations.

\begin{figure*}[tbp]
\centering
\includegraphics[width=0.8\textwidth]{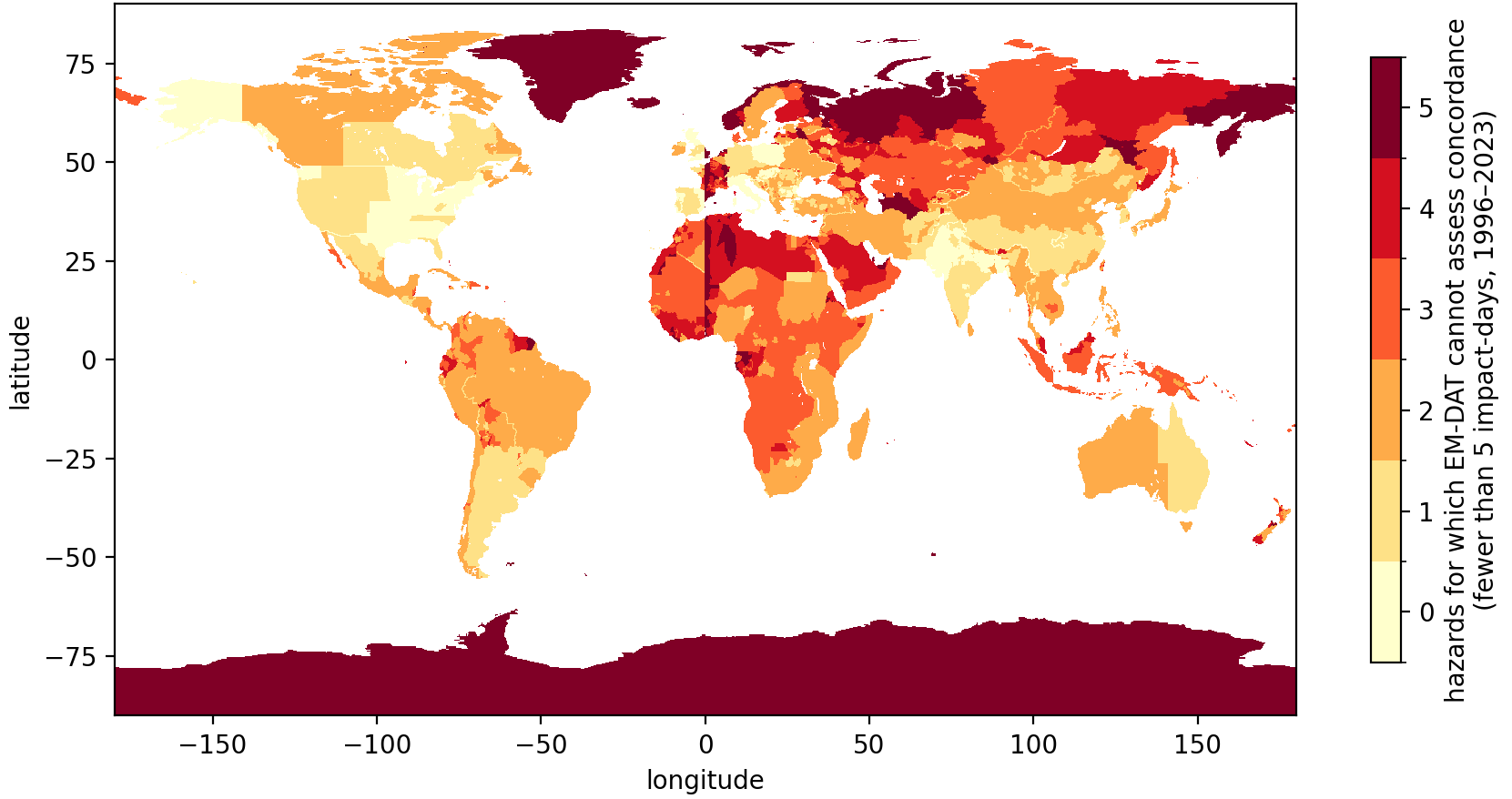}
\caption{\textbf{Where documented disaster records cannot evaluate a tail-event
dataset.} Per land cell, the number of hazards (0--5) for which EM-DAT
records fewer than five impact days over 1996--2023 and so cannot support a
local estimate (Sect.~\ref{sec:coverage}). Large contiguous regions,
concentrated in Africa, South America, and interior Asia, are uninformative
for three or more hazards at once; only 4.7\% of land is informative
for all five.}
\label{fig:coverage-gap}
\end{figure*}

EM-DAT is too sparse to assess this relationship across much of the world.
For heat, cold, wind, and drought, more than half of global land has too few
records for a local comparison. Heavy precipitation has denser flood reporting,
but 40.2\% of land is still unassessable (Fig.~\ref{fig:coverage-gap}).
Overall, 39.1\% of land falls below the screen for all five hazards. These gaps do
not generally indicate quieter weather: for four hazards, excluded areas have
similar or higher tail-event rates than areas that can be assessed.

The same gap affects 80.5\% of the world's land population: they live in a
place where the catalog cannot support a local estimate for at least one hazard
\citep{lloyd2017worldpop}. A benchmark based only on reported disasters would
therefore leave much of the inhabited world unassessed. TailWeather is
calculated in the same way wherever ERA5 provides land data, which makes it a
useful complement rather than a replacement for impact records.

\subsection{Tail labels overlap documented disasters}
\label{sec:capture}

\begin{figure*}[tbp]
\centering
\includegraphics[width=\textwidth]{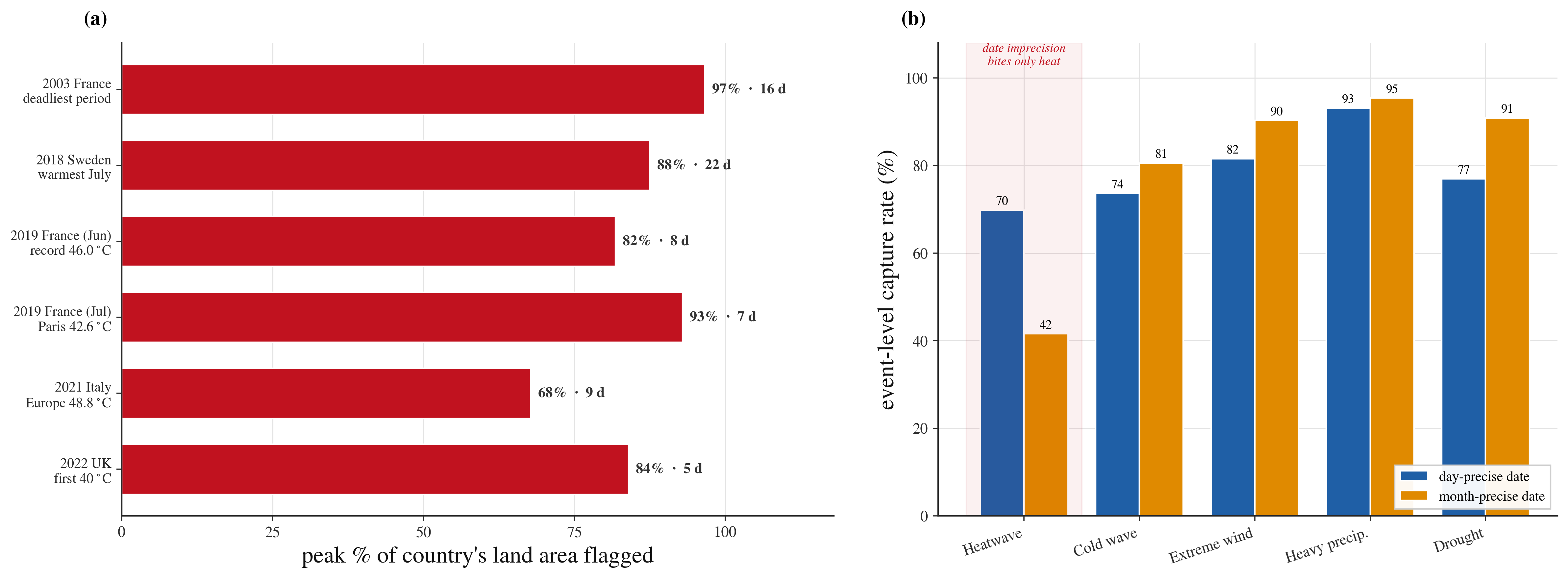}
\caption{Case studies and sensitivity to catalog date precision. (a) Six European heatwaves compared with TailWeather using externally reported dates and approximate national bounding boxes. Bars show peak flagged land-area fraction and flagged duration; the supplied analysis reports peak intensity 1.0 in each case. (b) Event-level capture stratified by EM-DAT date precision. Differences between groups indicate sensitivity to catalog construction and do not attribute all unmatched events to dating errors. }

\label{fig:case-study}
\end{figure*}

In this section, we ask the most direct question: do the labels match
with documented disasters? We first test the heatwave label against six
well-documented European heatwaves of the ERA5 era. All six match the
TailWeather label at the maximum intensity score, and their flagged durations
match the reported event lengths (Fig.~\ref{fig:case-study}a).

Across all EM-DAT records, we counted a disaster as captured when the
persistence-filtered label occurred anywhere within its reported area and
dates. Capture ranges from 55\% for heat to about 93\% for heavy precipitation
and drought, with cold and wind in between (Appendix~\ref{app:capture}). These raw rates do not measure detection skill: a large or long-lasting disaster footprint has more opportunities to overlap a tail event by chance. The illustrative chance comparison in Appendix~\ref{app:capture} shows why area, duration, and tail prevalence matter.

Heat has the lowest event-level capture. Records with month-level dates are captured less often than day-precise records (Fig.~\ref{fig:case-study}b), and wider matching windows increase capture. This shows sensitivity to catalog dating, but does not establish that all unmatched events result from date imprecision.

\subsection{Stricter tails are more often linked to recorded impacts}
\label{sec:refq}
\label{sec:monotone}

The event-level results ask whether a documented disaster overlaps any tail
event. We now sweep the tail threshold and compare individual cell-days. This
answers a different question: how does a stricter physical definition change
the balance between finding recorded impacts and flagging places without one?
The sweep is summarised by $q^{\ast}$, the threshold with the highest $F_1$
score (the harmonic mean of precision and recall). It is a summary of this comparison,
not a replacement for each hazard's label definition.

For heat, cold, and wind, $q^{\ast}$ is much stricter than the standard label
threshold. At $q=0.90$, a common diagnostic score threshold, recall is 15\% for
heat, 9.8\% for cold, and 13\% for wind (Table~\ref{tab:validation-tables}).
In this comparison, the selected thresholds trade lower recall for higher precision. For heavy precipitation and drought, the reported optimum lies below $q=0.90$. For drought, $q$ is a scaled SPI score, not a percentile; the released event definitions remain those in Appendix~\ref{app:defs}. This is why one
percentile cannot serve every hazard or purpose.

\label{sec:incommensurate}

Tail cell-days and recorded-impact cell-days also have very different base
rates. For example, at the standard threshold, heatwave tails cover
4.92\% of land cell-days, whereas recorded heat impacts cover only
0.104\%. Raw precision is therefore low even for a useful label. We
instead report \emph{enrichment}, or lift, $P(X\mid T_q)/P(X)$: how many times
more likely a recorded impact is on a tail cell-day than on a typical cell-day. This difference in base rates is not the whole explanation for low precision:
at $q=0.99$, heat tails are close in number to impact cells, but their observed
precision is still only 0.96\%. Thus, tail labels and recorded impacts are
related, but they are not interchangeable.
\begin{figure*}[tbp]
\centering
\includegraphics[width=\textwidth]{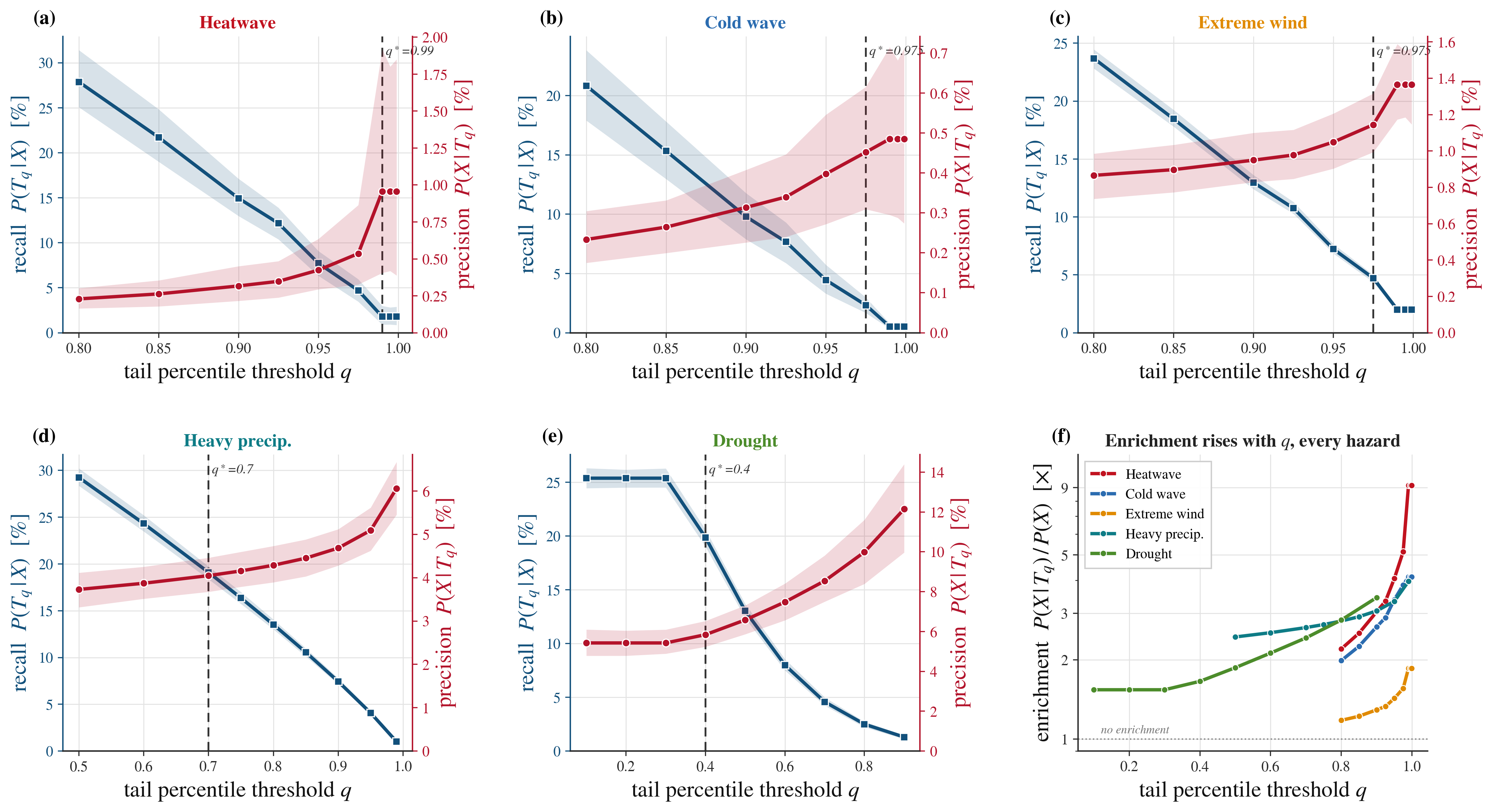}
\caption{Plots (a--e): recall $P(T_q\mid X)$ (blue, left axis) and precision
$P(X\mid T_q)$ (red, right axis) as functions of the tail threshold $q$ for each
hazard. The two conditionals are drawn on independent axes; the dashed line marks each hazard's F1-optimal $q^{\ast}$. Recall falls and precision rises monotonically in every plot. Plot (f) overlays the resulting enrichment
$P(X\mid T_q)/P(X)$ for all five hazards on a common logarithmic axis: it
increases monotonically with $q$ in every case, showing that the intensity score
carries impact information (Sect.~\ref{sec:monotone}).}
\label{fig:conditionals}
\end{figure*}
As the threshold becomes stricter, recall falls and precision rises
(Fig.~\ref{fig:conditionals}). More importantly, enrichment rises for all five
hazards: a stricter tail is more likely to coincide with a recorded impact.
The intensity score therefore contains useful impact-related information, even
though it is not itself a measure of harm.

\begin{table*}[tbp]
\centering
\caption{Catalog-overlap summary. (a) Cell-level recall $P(T_q\mid X)$ at the
F1-optimal $q^{\ast}$ and at the diagnostic score threshold $q=0.90$. (b)
Association between tail labels and recorded impacts, 1996--2023, global, land
only cells. The $q^{\ast}$ column names each hazard's own F1-optimal
threshold, but only Precision, Recall, and Lift are evaluated there; Tail(\%)
and Impact(\%) are instead each hazard's base rate at its own conventional
label threshold (Table~\ref{tab:overview}, e.g.\ P90 for heat), not at
$q^{\ast}$ -- heatwave's Tail(\%)~$=4.92$ matches Table~\ref{tab:overview}'s
overall event-day rate ($4.566$), not its extreme-tier ($I\ge0.99$) rate of
$0.648$. This mixes two thresholds in one row and is easy to misread; readers
comparing Tail/Impact against Precision/Recall/Lift should treat them as two
separate threshold conventions reported side by side, not as all six columns
sharing $q^{\ast}$.}
\label{tab:validation-tables}
\begin{minipage}[t]{0.3\textwidth}
\centering
\footnotesize
\textbf{(a) Recall at $q^{\ast}$ and P90}\\[3pt]
\resizebox{\linewidth}{!}{%
\begin{tabular}{@{}l S S S@{}}
\toprule
\textbf{Hazard} & {$q^{\ast}$} & {P90} & {$\times$} \\
\midrule
Heatwave      & 1.79  & 14.95 & 8.4 \\
Cold wave     & 2.33  & 9.79  & 4.2 \\
Extreme wind  & 4.71  & 12.97 & 2.8 \\
Heavy precip. & 19.10 & 7.44  & 0.4 \\
Drought       & 19.87 & 1.31  & 0.1 \\
\bottomrule
\end{tabular}}
\end{minipage}\hfill
\begin{minipage}[t]{0.5\textwidth}
\centering
\textbf{(b) Tail--impact association}\\[3pt]
\small
\setlength{\tabcolsep}{4.5pt}
\renewcommand{\arraystretch}{1.2}
\resizebox{\linewidth}{!}{%
\begin{tabular}{@{}l S S S S S S@{}}
\toprule
\textbf{Hazard} & {$q^{\ast}$} & {Tail} & {Impact} & {Precision} & {Recall} & {Lift} \\
 & & {(\%)} & {(\%)} & {(\%)} & {(\%)} & {($\times$)} \\
\midrule
Heatwave      & 0.99  & 4.92  & 0.104 & 0.96 & 1.79  & 9.2 \\
Cold wave     & 0.975 & 3.67  & 0.118 & 0.45 & 2.33  & 3.8 \\
Extreme wind  & 0.975 & 10.05 & 0.736 & 1.14 & 4.71  & 1.6 \\
Heavy precip. & 0.70  & 1.23  & 1.528 & 4.05 & 19.10 & 2.6 \\
Drought       & 0.40  & 3.67  & 3.530 & 5.84 & 19.87 & 1.7 \\
\bottomrule
\end{tabular}}
\end{minipage}
\end{table*}

\subsection{One threshold does not fit all hazards}
\label{sec:threshold}

\begin{figure*}[tbp]
\centering
\includegraphics[width=\textwidth]{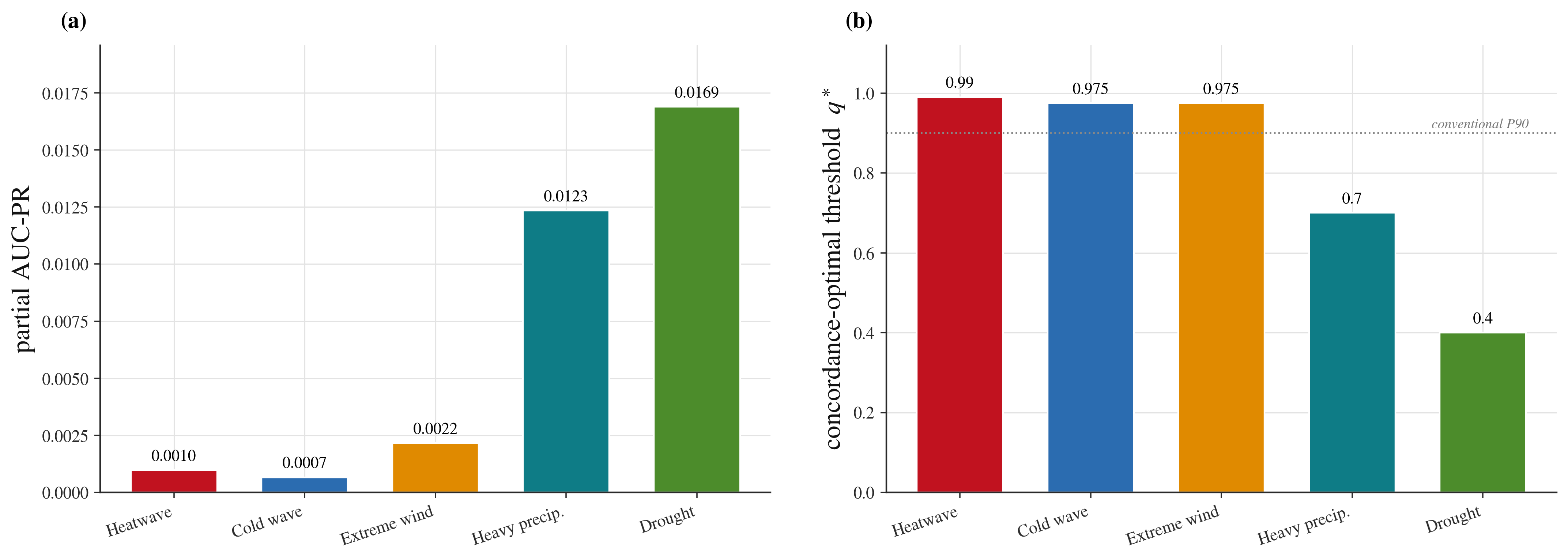}
\caption{\textbf{No universal ``right'' percentile.} (a) Partial area under the
precision--recall curve per hazard. This quantity is not comparable across
hazards, since it tracks each hazard's impact base rate more than its
predictability, and it is integrated only over the recall range the sweep spans
(Sect.~\ref{sec:assessment-caveats}). (b) F1-optimal score threshold $q^{\ast}$ for each hazard. The dotted line marks $q=0.90$; drought uses a scaled SPI score and should not be read on a common percentile scale.}
\label{fig:headline}
\end{figure*}

The reported $F_1$-maximizing threshold ranges from 0.99 for heat to 0.40 for drought (Fig.~\ref{fig:headline}). These values summarize this catalog comparison; they are not universal hazard thresholds. Users should choose thresholds for their target and the relative importance of missed events and unmatched detections. Differences in score scaling and precision also matter, as detailed in Appendix~\ref{sec:assessment-caveats}.

\subsection{The relationship differs by hazard and place}
\label{sec:hazard-differences}

The association is weakest for extreme wind. Six-hourly ERA5 fields on the \ang{0.25} grid do not resolve short-lived gust maxima. This scale mismatch is one plausible contributor to the weaker wind-impact association. The wind label is therefore a useful
physical target but an imperfect proxy for wind impacts.

\begin{figure*}[tbp]
\centering
\includegraphics[width=\textwidth]{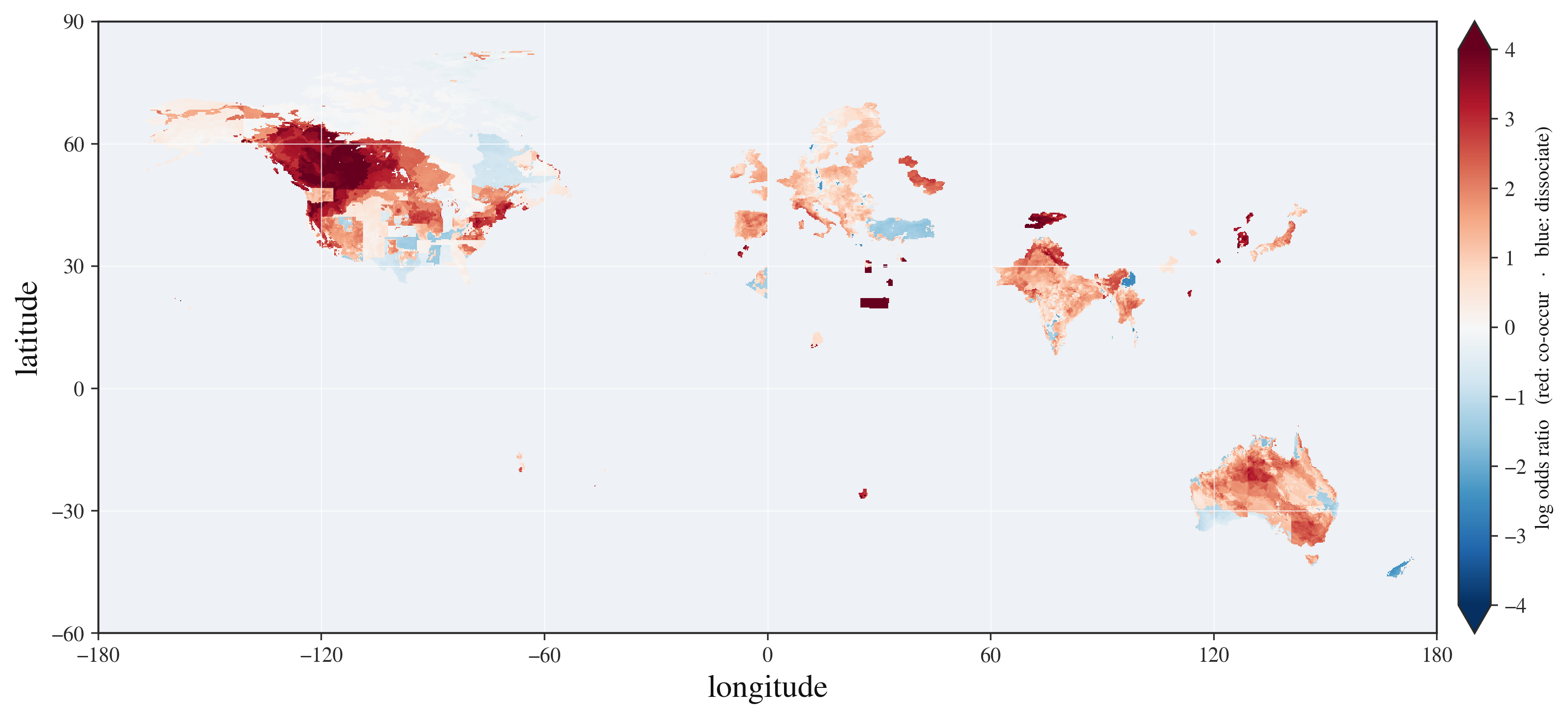}
\caption{\textbf{Heatwave tail--impact log-odds ratio at $q=0.90$.} Per-cell log
odds ratio between heatwave tail labels and recorded impacts. Cells
with fewer than five impact days are masked, and the diverging scale is clipped
symmetrically at $\pm4$ so that a
small number of saturated coastal cells (a rasterisation edge effect) do not
dominate the colour range. The association is positive nearly everywhere it can
be estimated; the blank areas over Africa and South America reflect absent
recorded impacts rather than absent tail events (Sect.~\ref{sec:spatial}).}
\label{fig:map-heat}
\end{figure*}

\begin{figure*}[tbp]
\centering
\includegraphics[width=\textwidth]{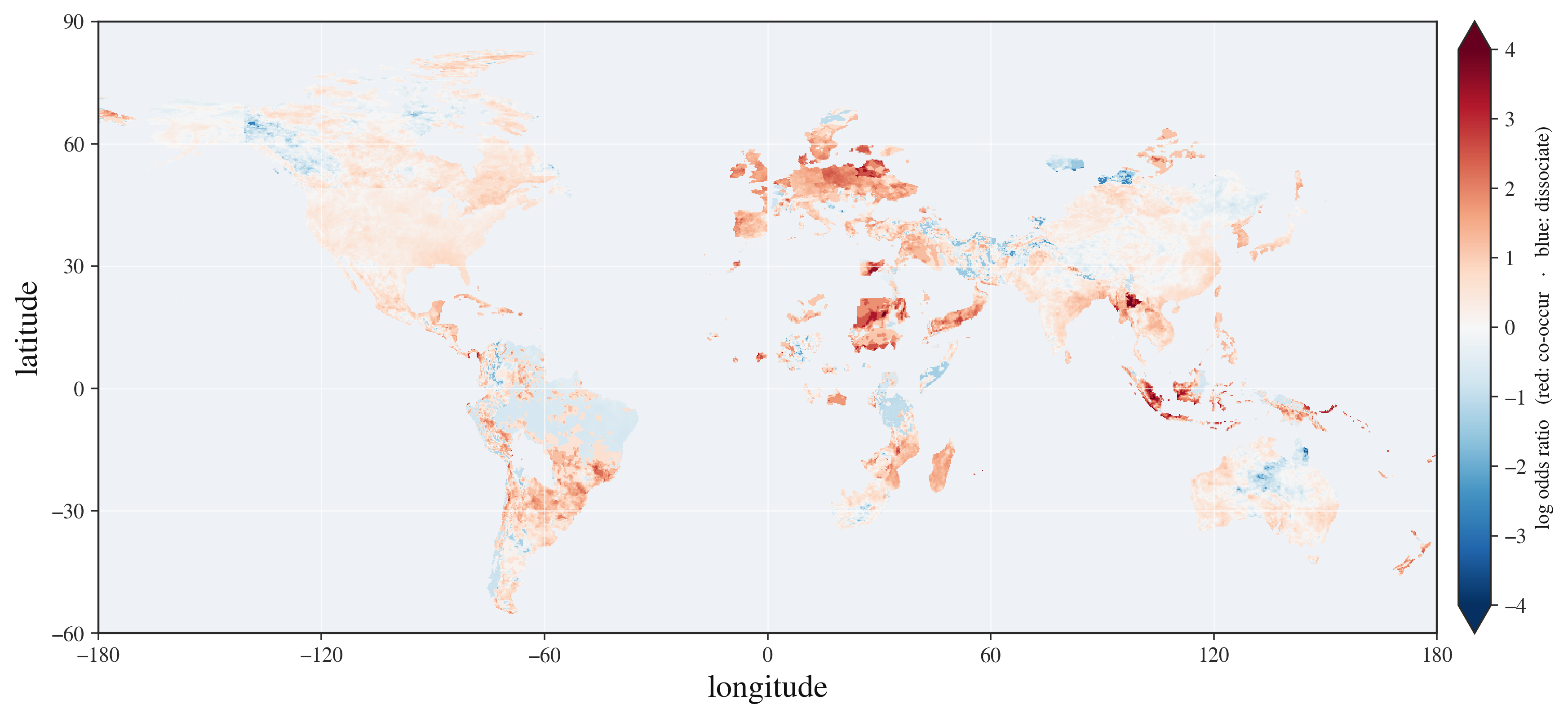}
\caption{\textbf{Extreme-wind tail--impact log-odds ratio at $q=0.90$.} As
Fig.~\ref{fig:map-heat}, but for extreme wind. The association is
visibly weaker and more spatially incoherent, consistent with the weaker pooled relationship between the wind-tail proxy and reported impacts
(Sect.~\ref{sec:hazard-differences}).}
\label{fig:map-wind}
\end{figure*}

\label{sec:spatial}
Figures~\ref{fig:map-heat} and \ref{fig:map-wind} map the local association
where at least five impact days are available. It is generally positive where
it can be estimated, but most of Africa and South America remain blank because
the record is too sparse, not because tail events are absent. Reporting follows
population, institutional capacity, and media attention. A disaster-only
benchmark would inherit this uneven geography; a local-climatology tail dataset
can evaluate the physical weather event consistently in both well-reported and
poorly reported regions.

\section{Compound tail events}
\label{sec:compound-concordance}

\begin{figure*}[tbp]
\centering
\includegraphics[width=0.8\textwidth]{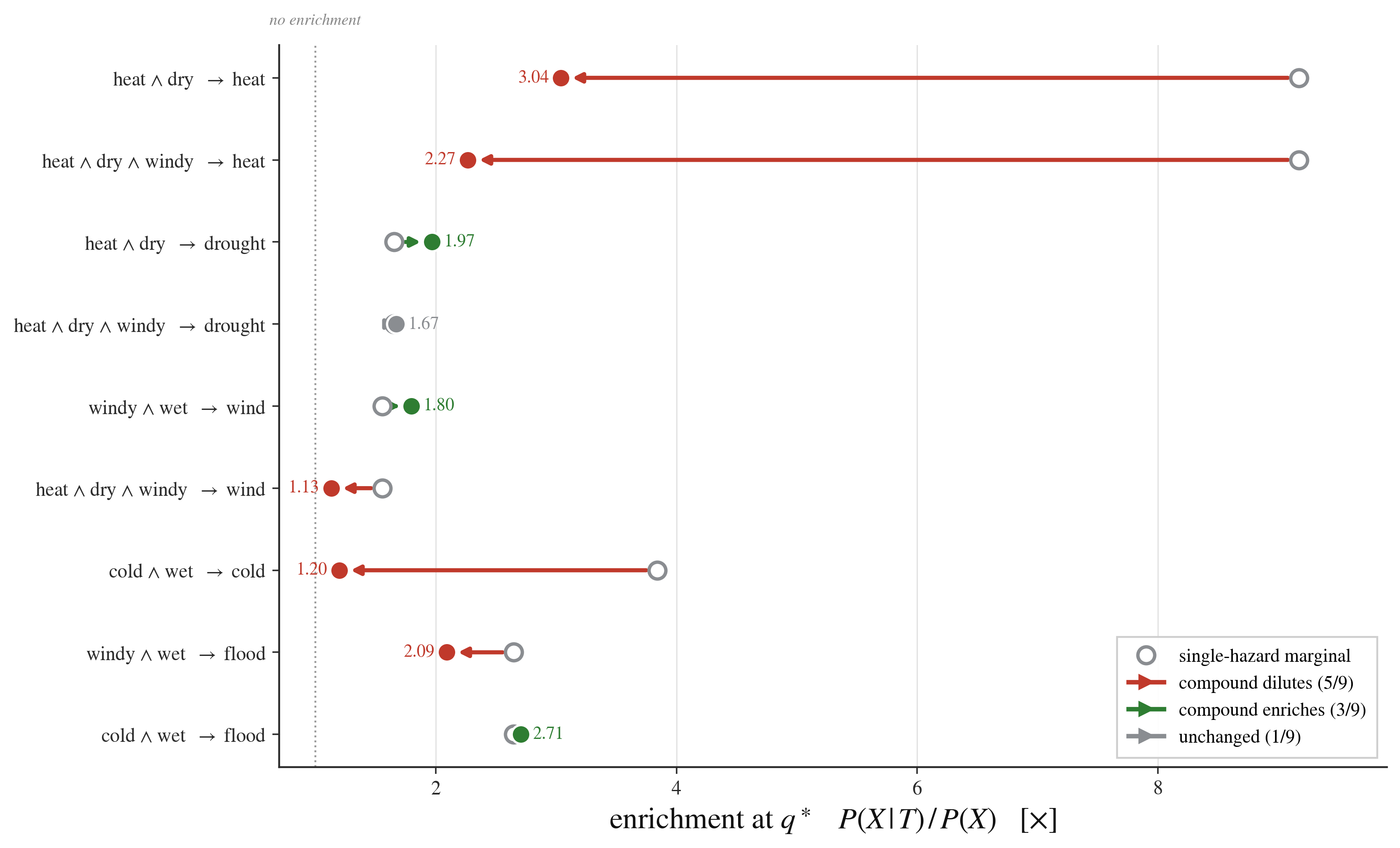}
\caption{\textbf{Compounding does not systematically sharpen the tail--impact
correspondence.} Each row is one compound label compared against the EM-DAT
impact record of one of its constituents (written
\textit{compound $\to$ constituent}). The hollow marker is the constituent single-hazard enrichment at its own $q^{\ast}$; the filled marker is the compound's enrichment at its own $q^{\ast}$; the arrow runs from the former to the latter. Red arrows (five of nine) show compounding \emph{diluting} the correspondence, green (three) \emph{strengthening} it, grey (one) leaving it unchanged.}
\label{fig:compound}
\end{figure*}

TailWeather also makes compound tails easy to derive. For a compound, we use
the lower of the constituent intensity scores: a hot--dry cell-day must be
both hot and dry at the selected threshold. We test hot--dry, hot--dry--windy,
windy--wet, and cold--wet labels against the documented impacts of each
constituent. Compounding is not automatically more informative. In five of nine comparisons it weakens the association with recorded impacts, in three it strengthens it, and in one it makes no difference (Fig.~\ref{fig:compound}). For example, adding drought and wind to heat lowers the association with recorded heat impacts, consistent with some recorded heat disasters lacking all three conditions. A hot--dry label is, however, slightly more associated with recorded drought impacts than drought alone. Compound labels should therefore be chosen for the process of interest. These comparisons use each label's own fitted threshold, so they do not isolate the effect of compounding alone.

\section{Example application: AI model training with TailWeather}
\label{sec:finetune}

TailWeather can also be used as a dense training signal. We test this idea with
a deliberately limited experiment: a small trainable sharpening head is added
to frozen FourCastNet (FCN) forecasts. The head is trained to give greater
weight to cells and days with higher TailWeather intensity scores. It corrects
the final forecast only; it does not change the FCN rollout. Implementation
details are given in Appendix~\ref{app:sharpening}.

\begin{figure*}[tbp]
\centering
\includegraphics[width=0.75\textwidth]{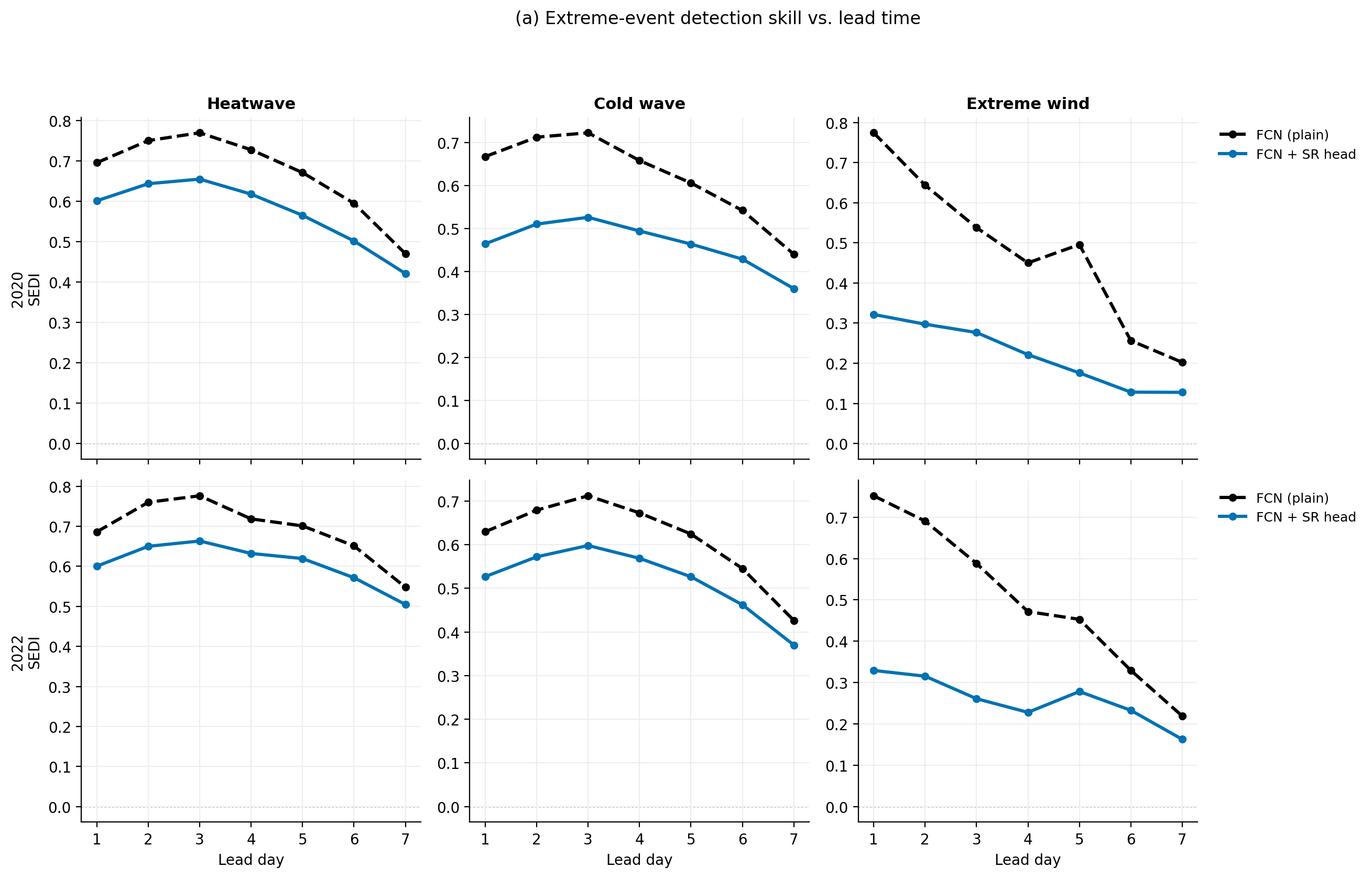}
\includegraphics[width=0.75\textwidth]{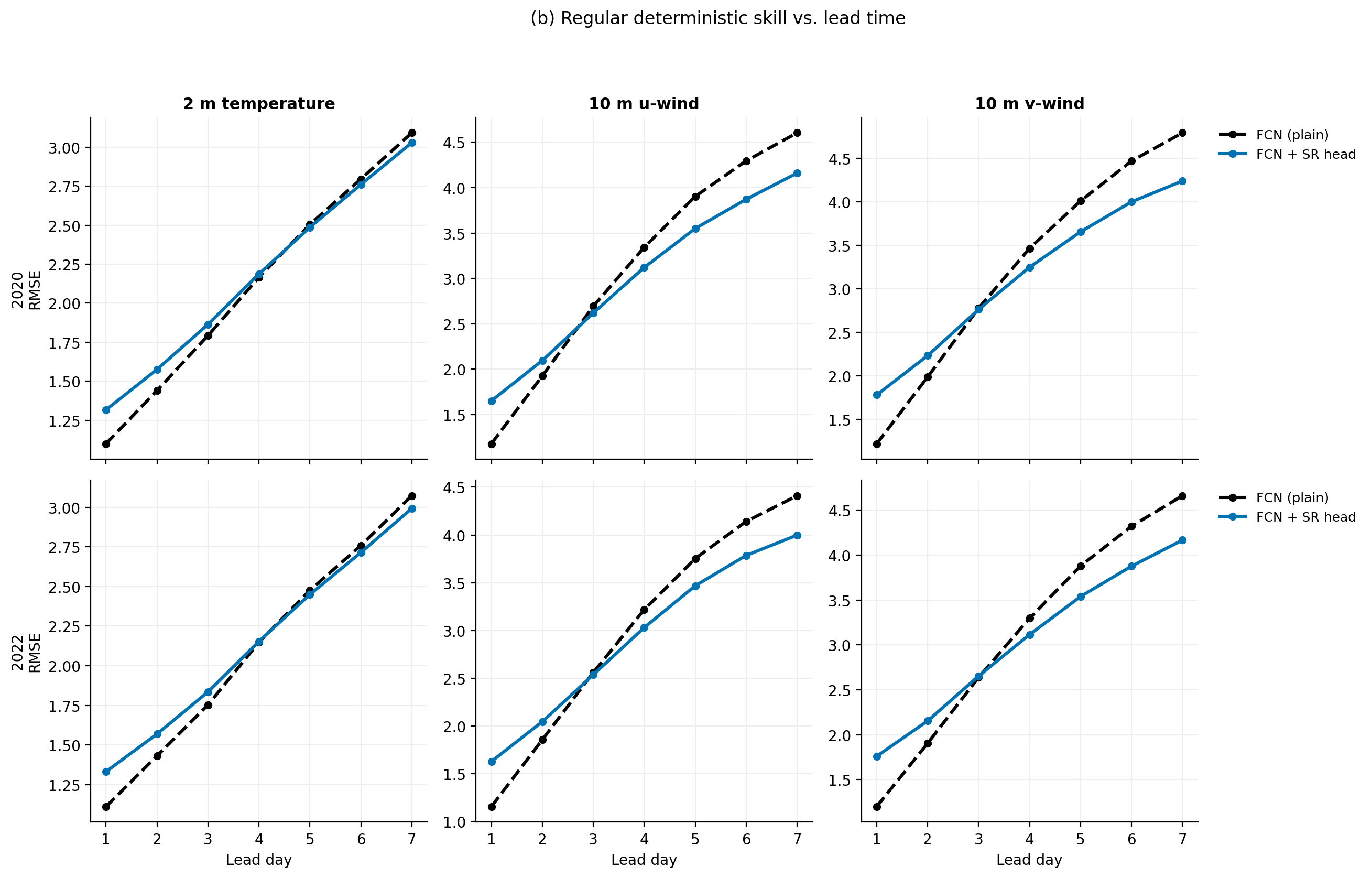}
\caption{\textbf{A tail-aware correction improves regular skill but degrades extreme-event detection.} Frozen FourCastNet (FCN) forecasts with (solid) and without (dashed) a trainable sharpening head applied as post-processing. (a) Extreme-event detection skill (SEDI, the higher the better) falls at every lead day. (b) Regular deterministic skill (latitude-weighted RMSE) tells the opposite story at longer lead times: the correction lowers 10~m wind RMSE by 30--55\,\% by day~7, and 2~m temperature RMSE more modestly.}
\label{fig:sr-head}
\end{figure*}

The experiment shows both the value and the limit of this signal. The correction
reduces 10~m wind RMSE by 30--55\% at day~7, so it can improve an ordinary
pointwise forecast. Yet extreme-event detection becomes worse at every lead
time, hazard, and test year. We measure detection with the Symmetric Extremal
Dependence Index (SEDI), a measure developed for rare binary events \citep{ferro2011sedi}. The largest decline is 0.45 for
extreme wind at day~1 (Fig.~\ref{fig:sr-head}). This experiment shows that improving pointwise accuracy does not necessarily improve event detection. It does not establish the mechanism behind that trade-off.

\section{Discussion}
\label{sec:discussion}

TailWeather and disaster catalogs serve different purposes. TailWeather labels
weather that is rare for a location; a disaster record describes reported harm,
which also depends on exposure, vulnerability, and reporting. The two are
related: stricter tails are more often associated with recorded impacts. But
they should not be treated as interchangeable, especially where disaster
reporting is sparse. TailWeather provides a global physical target; impact
catalogs remain necessary when the question is societal consequence.

The dataset makes an otherwise hidden forecasting question measurable. AI
systems can perform well on average while failing to detect some tail events,
especially extreme wind. It also supplies a training signal, but the sharpening
experiment shows that this signal alone is not enough: the correction improves
pointwise accuracy while reducing event-detection skill. Thresholds and compound
labels should likewise be chosen for the hazard and question, not applied as
universal rules.

Every label inherits ERA5's biases and the limitations of the \ang{0.25}, \SI{6}{\hour} input sampling \citep{hersbach2020era5}. The dataset is not suitable for convective
gusts, tornadoes, hail, or downbursts, and the heavy-precipitation and drought
labels have their own temporal limitations. Thresholds use a fixed 1991--2020
reference climate, so they describe rarity relative to that baseline rather
than a moving climate. Source uncertainty, threshold estimation, and incomplete impact reporting all limit interpretation. Detailed processing and comparison caveats are given in Appendices~\ref{app:processing} and \ref{sec:assessment-caveats}.

\section{Code and data availability}
\label{sec:availability}

\dataavailability{TailWeather is available at \url{https://huggingface.co/datasets/zhisong-liu/TailWeather}, with illustrative subsets and a demonstration at \url{https://huggingface.co/spaces/zhisong-liu/TailWeather}. The repository declares the Apache License 2.0. Source ERA5 data are available from the Copernicus Climate Data Store and WeatherBench2 \citep{hersbach2020era5,rasp2024weatherbench2}. Impact records remain subject to their maintainers' access terms.}

\codeavailability{The dataset repository contains label files and processing logs. A versioned archive of the generation and analysis code is needed to make the workflow reproducible.}

\conclusions
\label{sec:conclusions}

TailWeather provides global land labels for five climatological hazards on a \ang{0.25} grid, covering 1981--2022 and part of January 2023. Its severity tiers and intensity scores provide a dense physical target for evaluating and developing weather models.

The labels overlap many documented disasters and stricter tails are more often
associated with recorded impacts, but rarity is not harm. The relationship
changes by hazard, compound labels are not automatically better, and disaster
records cannot evaluate much of the world. TailWeather consequently complements
rather than replaces impact data. Its AI applications show why this distinction
matters: strong average forecast skill can conceal weak extreme-event detection,
and a tail-aware correction can improve RMSE while worsening that detection.

\appendix
\section{Event definitions}
\label{app:defs}

Each hazard is flagged by one reproducible percentile-and-persistence rule
applied to a single ERA5 variable against the local 1991--2020 climatology;
Sect.~\ref{sec:source} summarizes them and Table~\ref{tab:overview} lists the
parameters. The complete definitions follow.

\subsection{Heatwave}
\label{sec:hw}

We follow the percentile-and-persistence definition of
\citet{perkins2013heatwave}. Let $\mathrm{TX}(c,d)$ denote the daily maximum
\SI{2}{\meter} temperature at land cell $c$ on date $d$ with day-of-year
$j=\mathrm{doy}(d)$. A seasonally varying threshold is the local 90th percentile
of the reference climatology over a centered $\pm15$-day window,
\begin{equation}
T_{90}(c,j) = P_{90}\!\left\{\, \mathrm{TX}(c,d') : d'\in\text{ref. period},\;
|\mathrm{doy}(d')-j|\le 15\ (\text{circular})\,\right\}.
\end{equation}
A day is an exceedance if $E(c,d)=\mathbf{1}[\mathrm{TX}(c,d)>T_{90}(c,j)]$, and
a heatwave is flagged where exceedances persist for at least three consecutive
days,
\begin{equation}
\mathrm{HW}(c,d)=\mathbf{1}\big[\exists\text{ a run of }\ge 3\text{ consecutive
days with }E=1\text{ that includes }d\big].
\end{equation}
The intensity score $I(c,d)$ is the empirical percentile rank of
$\mathrm{TX}(c,d)$ in the local climatological cumulative distribution function
(CDF); severity tiers bin it at
$0.90/0.95/0.99$.
\emph{Parameters:} $P_{90}$, $\pm15$-day window, $\ge 3$ days, reference
1991--2020.

\subsection{Cold wave}
\label{sec:cw}

The cold wave is the lower-tail mirror of the heatwave definition. Let
$\mathrm{TN}(c,d)$ be the daily minimum \SI{2}{\meter} temperature. Using the
local 10th percentile over the same $\pm15$-day seasonal window,
\begin{equation}
T_{10}(c,j) = P_{10}\!\left\{\, \mathrm{TN}(c,d') : d'\in\text{ref. period},\;
|\mathrm{doy}(d')-j|\le 15\,\right\},
\end{equation}
a day is an exceedance if $E(c,d)=\mathbf{1}[\mathrm{TN}(c,d)<T_{10}(c,j)]$ and a
cold wave requires a run of at least three consecutive exceedance days. The
cold-tail intensity score is $I(c,d)=1-\text{(percentile rank)}$, so a value
colder than the entire reference distribution scores $\approx 1$; tiers use the
same $0.90/0.95/0.99$ bins. The three-day minimum is the TailWeather choice and should not be identified with the six-day ETCCDI cold-spell duration index. \emph{Parameters:} $P_{10}$, $\pm15$-day window, $\ge 3$ days,
reference 1991--2020.

\subsection{Heavy precipitation}
\label{sec:precip}

We adopt the wet-spell taxonomy of \citet{zolina2010precip}, combining the
95th-percentile wet-day threshold (R95p) with wet-spell structure. Let $P(c,d)$ be daily total
precipitation. A \emph{wet day} has $P\ge\SI{1}{\milli\meter}$; a \emph{wet
spell} is a run of $\ge 2$ consecutive wet days. The heavy-precipitation
threshold is the 95th percentile over the \emph{wet-day} distribution only,
\begin{equation}
R_{95}(c)=P_{95}\!\left\{\,P(c,d') : d'\in\text{ref. period},\;
P(c,d')\ge\SI{1}{\milli\meter}\,\right\},
\end{equation}
so that $\mathrm{heavy}(c,d)=\mathbf{1}[P(c,d)\ge\SI{1}{\milli\meter}\wedge
P(c,d)>R_{95}(c)]$. The structural (Zolina) diagnostic flags a heavy day
embedded in an anomalously long wet spell: with $L(c,d)$ the length of the wet
spell containing $d$ and $S_{75}(c)$ the 75th percentile of reference wet-spell
lengths,
\begin{equation}
\mathrm{heavy\_long}(c,d)=\mathbf{1}\big[\mathrm{heavy}(c,d)=1 \wedge
L(c,d)\ge S_{75}(c)\big].
\end{equation}
The intensity score is the wet-day percentile rank of $P(c,d)$.
Because the R95p criterion already places every flagged day at or above the
95th wet-day percentile, the moderate tier $[0.90,0.95)$ is empty by
construction: heavy precipitation has no $s=1$ events
(Table~\ref{tab:overview}). This is a property of the definition, not a
processing error. \emph{Parameters:} wet-day $=\SI{1}{\milli\meter}$, $R_{95}$
over wet days, $P_{75}$ spell length, minimum spell $2$ days, reference
1991--2020.

In the inspected 2023 release, \texttt{precip\_severity}>0 matches \texttt{heavy\_precip}>0, while \texttt{heavy\_long\_spell} is a separate field. A heavy-precipitation label therefore does not require membership in a multi-day spell in this release.

\subsection{Extreme wind}
\label{sec:wind}

Let $W(c,d)=\max_{t\in d}\sqrt{u_{10}(c,t)^2+v_{10}(c,t)^2}$ be the daily
maximum sustained 10 m wind speed among the available six-hourly samples. The
labelling code will prefer the ERA5 instantaneous wind gust field
(\texttt{i10fg}, a 3-second peak) where present, since it is the more
impact-relevant, WMO-defined quantity, but no gust field was present in the
ERA5 archive used to build the released dataset, so every year's label
(verified in the \texttt{wind\_2020.nc} \texttt{source} attribute and variable
name, both stating sustained wind speed) uses the sustained-speed fallback
above, not gust. This must not be confused with a peak-gust product. With the
local 98th
percentile over the $\pm15$-day window,
\begin{equation}
W_{98}(c,j)=P_{98}\!\left\{\,W(c,d') : d'\in\text{ref. period},\;
|\mathrm{doy}(d')-j|\le 15\,\right\},
\end{equation}
a day is flagged if $W(c,d)>W_{98}(c,j)$. Because wind extremes are transient
synoptic storms lasting hours to about a day, the minimum duration is one day.
Severity tiers use speed boundaries motivated by the Beaufort scale: a flagged day is at least
moderate, escalating to severe at $\SI{20.8}{\meter\per\second}$ (Beaufort
forces 9--10) and extreme at $\SI{28.5}{\meter\per\second}$ (force 11 and above). ERA5 at \ang{0.25}
and \SI{6}{\hour} resolves synoptic-scale wind extremes but not convective
gusts; this is the reanalysis-derivable storm analogue, and
Sect.~\ref{sec:hazard-differences} quantifies what that proxy costs in
correspondence to recorded wind damage. \emph{Parameters:} $P_{98}$,
$\pm15$-day window, $\ge 1$ day, Beaufort tiers at
$20.8/\SI{28.5}{\meter\per\second}$, reference 1991--2020.

\subsection{Meteorological drought (SPI-3)}
\label{sec:drought}

Drought is defined by the Standardized Precipitation Index (SPI) at a 3-month
accumulation \citep{mckee1993spi}, standardized through a fitted gamma
distribution \citep{lloydhughes2002drought}. This is the only \emph{parametric}
tail in the dataset. Daily precipitation is aggregated to monthly totals and a
3-month rolling sum is formed, $P_3(c,m)=\sum_{k=0}^{2}P_{\text{month}}(c,m-k)$.
For each cell and target month a mixed distribution is fitted to the
reference-period accumulations to accommodate exact zeros,
\begin{equation}
H(x)=q+(1-q)\,G(x;\alpha,\beta),
\end{equation}
where $q=\Pr(P_3=0)$ and $G(\cdot;\alpha,\beta)$ is the gamma CDF fitted by
maximum likelihood to the non-zero values. The index is the standard-normal
quantile of the cumulative probability,
\begin{equation}
\mathrm{SPI}(c,m)=\Phi^{-1}\!\big(H(P_3(c,m))\big),
\end{equation}
Negative SPI values indicate drier-than-reference conditions. Approximate standard normality depends on the fitted distribution and is limited by the discrete probability mass at zero. A cell is in drought
when $\mathrm{SPI}\le-1.0$, with tiers at $-1.0/-1.5/-2.0$ following
\citet{mckee1993spi}. SPI-3 is inherently monthly; a daily-broadcast variant
replicates each month's label across its days. Broadcasting does not add temporal information. Repeating a monthly score on each day creates within-month dependence but does not, by itself, limit the range of score thresholds that can be evaluated. A complete month's precipitation also uses information unavailable earlier in that month; retrospective daily broadcast must not be used as a contemporaneous predictor.
For the inspected drought release, a score threshold $q=0.40$ corresponds to approximately $\mathrm{SPI}\le-1.2$ on finite-score cells; $q=0.90$ corresponds to $\mathrm{SPI}\le-2.7$. Neither is the 40th or 90th climatological percentile. Values below the default drought threshold cannot be reconstructed from the drought score alone because it is missing on non-drought cells; the stored \texttt{spi3} field is needed.


\section{Dataset characteristics}
\label{app:characteristics}

\begin{figure*}[t]
\centering
\includegraphics[width=0.9\textwidth]{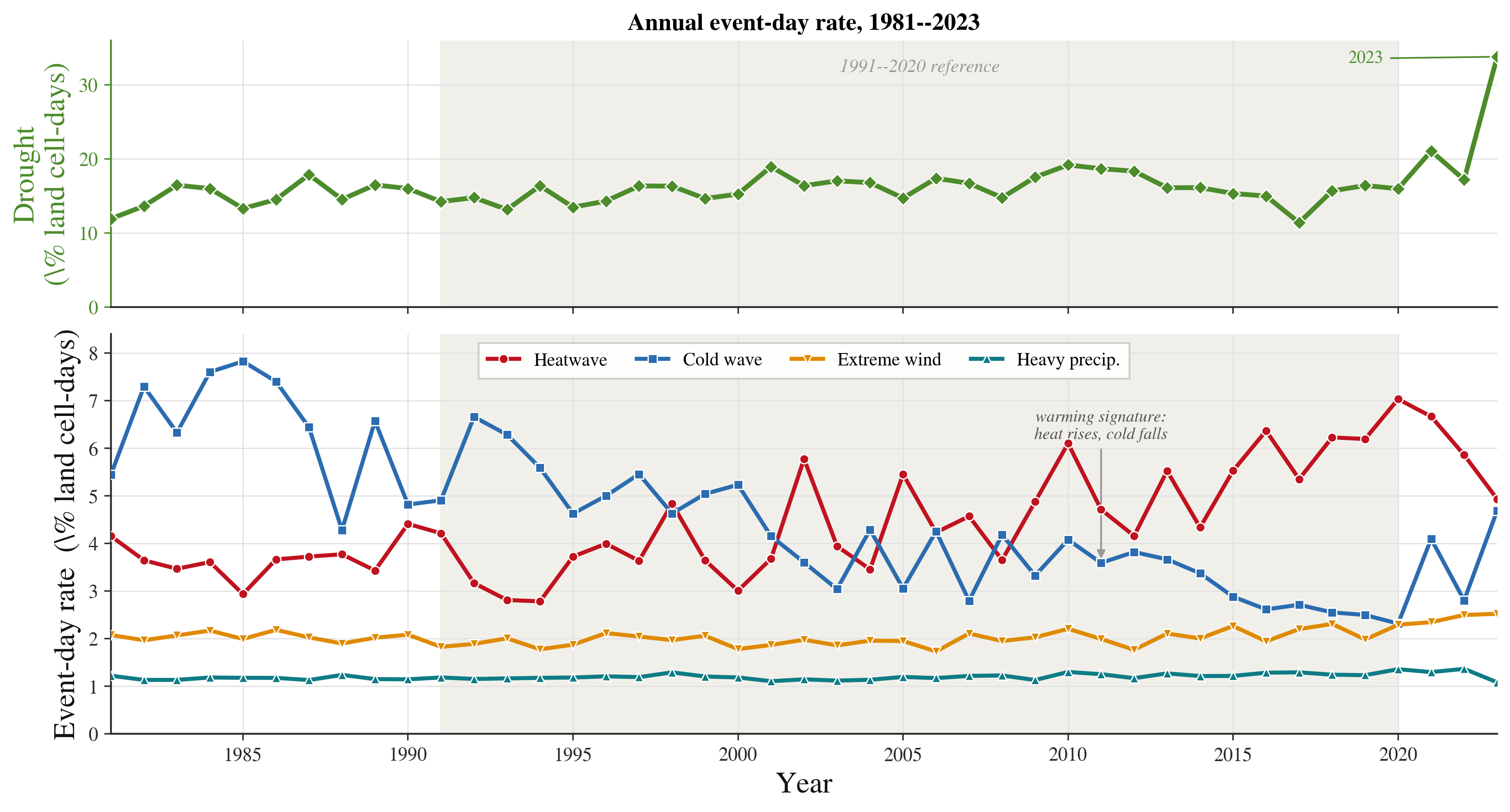}
\caption{Annual event-day rate for each hazard, 1981--2022, plus a partial
2023, as a percentage of land cell-days, with the 1991--2020 climatological reference period shaded. Drought (top) is shown on its own scale because it is four to eight times more prevalent than the other four hazards (bottom). Because thresholds are held fixed at the reference-period climatology, these curves show how each hazard's frequency has evolved relative to a stationary baseline: the heatwave and cold-wave series cross as heat rises and cold falls, the expected warming signature. The apparent drought value of 33.8\,\% at the 2023 endpoint is
not an annual rate: it is the single partial-January 2023.}
\label{fig:frequency}
\end{figure*}

\begin{figure}[t]
\centering
\includegraphics[width=\linewidth]{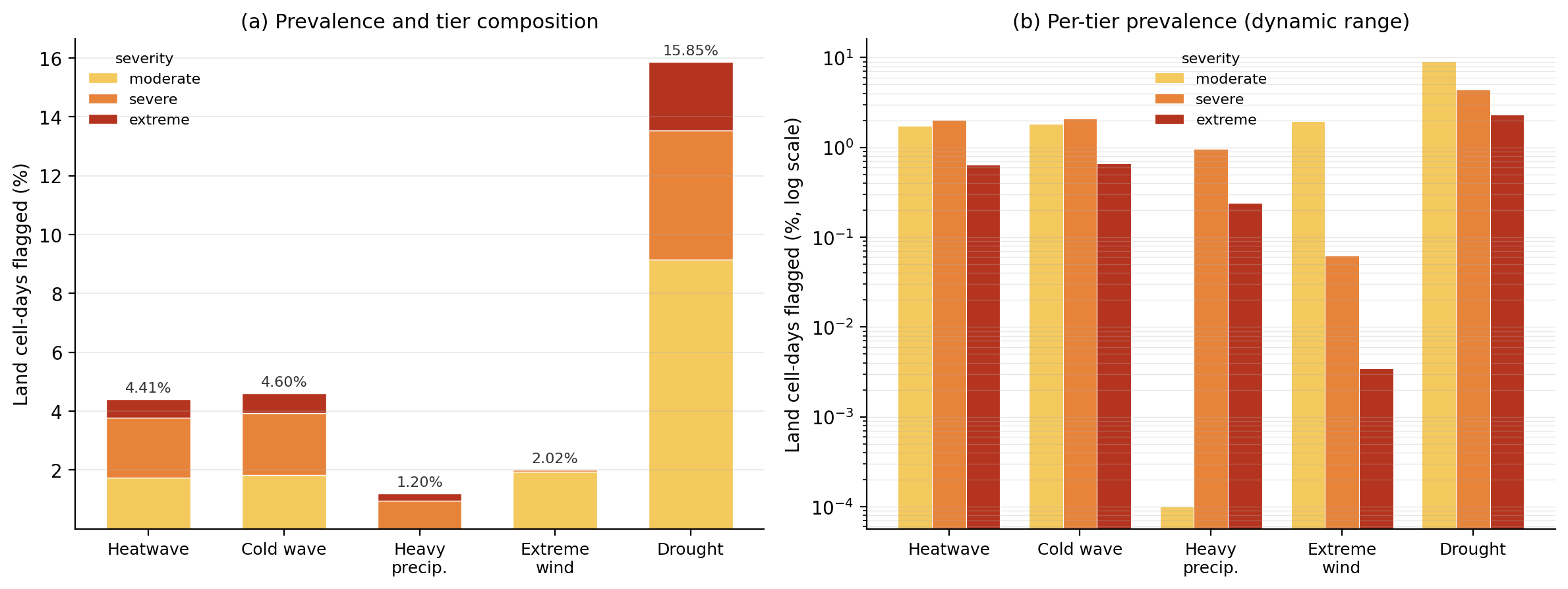}
\caption{Distribution of severity tiers within each hazard. Heavy precipitation
has no moderate tier by construction (Sect.~\ref{sec:precip}), and extreme
wind's Beaufort-anchored tiers place almost all flagged days in the moderate
band because the absolute \SI{20.8}{\meter\per\second} escalation threshold is
rarely met.}
\label{fig:severity}
\end{figure}

\begin{figure*}[t]
\centering
\includegraphics[width=\textwidth]{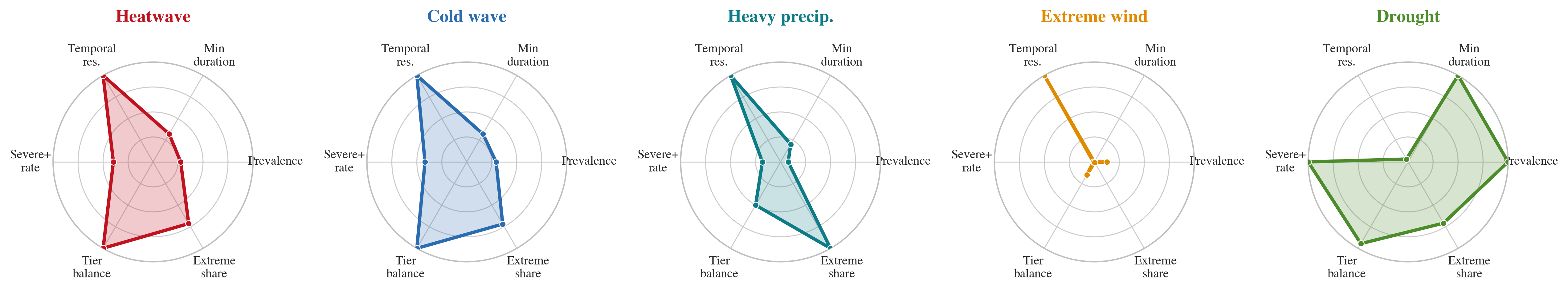}
\caption{Comparative profile of the five hazards across six dataset properties:
prevalence (event-day rate), minimum duration, temporal resolution, the
severe-or-extreme event rate, tier balance (the normalised entropy of the
moderate/severe/extreme mix), and extreme-tier share. Each axis is normalised to
its cross-hazard maximum. Drought is prevalent, long-duration, and coarse in
time, whereas extreme wind is rare, short, and strongly concentrated in its
lowest severity tier.}
\label{fig:radars}
\end{figure*}

This appendix summarizes the descriptive statistics of the labeled dataset:
annual frequency (Fig.~\ref{fig:frequency}), severity-tier distributions
(Fig.~\ref{fig:severity}), and a comparative per-hazard profile
(Fig.~\ref{fig:radars}).

The fixed 1991--2020 reference allows frequencies in different years to be compared against the same baseline. The supplied summaries report an increase in heatwave frequency from 3.7\,\% (1981--1990) to 5.9\,\% (2014--2023), and a decline in cold-wave frequency from 6.4\,\% to 3.1\,\%. These contrasts are consistent with a warming climate. Recomputing the recent-decade endpoint over nine complete years (2014--2022) instead of ten (2014--2023, including the partial January value) changes heat from 5.85\,\% to 5.95\,\% and cold from 3.05\,\% to 2.87\,\%: both conclusions are robust to whether the partial 2023 point is included. The supplied drought series reaches 33.8\,\% in 2023. Inspection reproduces approximately 33.81\,\% from the single January record (118968 drought cells among 351848 cells with finite SPI), so this value cannot be described as a full-year rate. Its input accumulation also needs checking for a partial January total.

A fixed reference does not, by itself, explain an anomalous terminal value or establish that it is an artifact. Such a value requires checking input completeness, precipitation accumulation, fitting, and spatial contributions. The stored scores quantify rarity relative to the original baseline. A moving-reference analysis would require refitting the climatology from the underlying meteorological fields.

\section{Processing details}
\label{app:processing}

This appendix records the construction details needed to reproduce the labels;
the per-hazard rules themselves are in Appendix~\ref{app:defs}.

\noindent\textbf{Terminal coverage and sampling.}
The cited WeatherBench2 archive ends at 18:00 UTC on 10 January 2023 \citep{wb2guide}. Inspection of the global 2023 label files finds 1--10 January for heat, cold, and wind; 1--9 January for precipitation; and one January coordinate for monthly drought. These partial periods cannot be treated as a full year. Daily extrema based on six-hourly samples differ from complete hourly extrema. Monthly precipitation and SPI require complete accumulation windows; incomplete terminal months should be excluded.

\noindent\textbf{Climatological reference distribution.}
For temperature and wind, the seasonal reference sample pools values within a circular $\pm15$-day window over 1991--2020, giving approximately $30\times31=930$ values before accounting for missing data and leap-day conventions. The precipitation equation in Sect.~\ref{sec:precip} instead pools wet days across the full reference period and does not include a seasonal window. SPI is fitted separately by calendar month. Pooling increases sample size but does not make consecutive weather observations independent.

\noindent\textbf{Circular day-of-year window.}
The $\pm15$-day window wraps around the year boundary: the reference sample for
early-January target days includes late-December reference values, and vice
versa, so thresholds are continuous across 31 December--1 January rather than
truncated at the calendar boundary.

\noindent\textbf{Leap days.}
Day-of-year is the calendar's own \texttt{dayofyear} (1--365, or 1--366 in a
leap year), pooled across reference years without further adjustment. Because
about a quarter of the 1991--2020 reference years are leap years, this shifts
the nominal day-of-year of every date from 1~March onward by one day in those
years relative to non-leap years, not only 29~February itself: the reference
pool for a given target day mixes calendar dates that are up to a day apart in
about one reference year in four. The seasonal window may reduce sensitivity to this offset, but its effect has not been quantified. A comparison with a consistent calendar-date convention is needed before claiming that the difference is negligible. The implementation should also specify how circular distance treats day 366.

\noindent\textbf{Percentile estimator.}
Threshold percentiles are read from the sorted reference sample by linear
interpolation between the two nearest order statistics (\texttt{numpy}'s
default \texttt{percentile} convention, used unmodified throughout the
labelling code) rather than a nearest-rank rule. Rank spacing is of order $1/n$ for $n$ reference values, but the difference in physical threshold values depends on the spacing of the relevant order statistics.

\noindent\textbf{Intensity-score precision and far-tail saturation.}
The precision of stored scores and the resolution of the reference distribution are distinct. An earlier inspection of only the truncated 2023 files (Sect.~\ref{app:processing}) suggested a coarser, $k/99$-level grid, but this does not hold for the dataset as a whole: the pooled 43-year percentile tables (\texttt{label\_summary.json}) give the 50th/90th/95th/99th score percentile for all four daily hazards, and all sixteen values are exact integer multiples of $0.0005$ (e.g.\ heat P50~$=0.4845=969\times0.0005$) and none are close to a multiple of $1/99\approx0.0101$. The stored score grid is therefore $0.0005$, as originally reported; the coarser spacing apparent in the 2023 file is likely an artifact of its truncation to a handful of days rather than a property of the score encoding itself. Increasing numerical storage precision alone would not resolve the far tail (below), which is a property of the empirical rank, not of storage.

The monthly drought field differs: its finite values equal $\min(1,-\mathrm{SPI}/3)$ on drought cells, and it is missing elsewhere. This yields a minimum finite score near $1/3$ and explains why thresholds 0.1, 0.2, and 0.3 select the same finite-score set. Monthly broadcasting adds temporal dependence but is not the cause of this threshold plateau.

\section{Tail--impact comparison details}
\label{app:concordance-detail}

For valid comparison cell-days, define $a=\sum T_qX$, $b=\sum T_q(1-X)$, $c=\sum(1-T_q)X$, and $d=\sum(1-T_q)(1-X)$. Then
\begin{equation}
\mathrm{precision}=\frac{a}{a+b},\qquad
\mathrm{recall}=\frac{a}{a+c},\qquad
F_1=\frac{2a}{2a+b+c},\qquad
\mathrm{lift}=\frac{a/(a+b)}{(a+c)/(a+b+c+d)}.
\end{equation}
The same valid mask and weights must be used in all terms. Per-cell log odds ratios are $\log(ad/bc)$ where defined; zero counts require an explicitly stated correction or masking convention. Missing catalog records cannot be interpreted as verified absence of impact.

The draft reports 500 bootstrap resamples of 15-day blocks. To retain spatial dependence, temporal blocks should resample the full spatial field together. A 15-day block does not necessarily represent uncertainty for monthly SPI, overlapping three-month accumulations, or long disaster records. Confidence intervals from this procedure also omit uncertainty in source reanalysis, threshold estimation, geolocation, and catalog inclusion.

\section{Sharpening-head implementation}
\label{app:sharpening}

Starting from pre-trained FourCastNet \citep{fcn}, we froze the backbone and
attached a shallow residual head \citep{resnet}. The head reads the frozen
forecast of 2~m temperature and the two 10~m wind components, proposes a
post-processing correction, and predicts an attention map trained against the
TailWeather intensity score \citep{attention}. Losses give larger weight to
cells and days with higher intensity scores. The corrected forecast is not fed
back into the autoregressive FCN state. To expose the head to errors that grow through a rollout, training used increasing rollout lengths of 1, 3, 5, and 7 days. The experiment is intended as a simple test of whether TailWeather can supply useful supervision, not as a comparison of training methods.

\section{Limits of the tail--impact comparison}
\label{sec:assessment-caveats}

\noindent\textbf{Footprint and reporting uncertainty.}
Administrative polygons may substantially exceed the physical area affected, particularly for country-level records after the end of GDIS coverage. Enlarging a footprint increases the opportunity for an any-overlap match, but the effects on pooled precision, recall, and odds ratios are not universal upper or lower bounds. The comparison should be stratified by geometry, event period, and date precision.

\noindent\textbf{Cell-day and event-level estimands.}
EM-DAT applies disaster-inclusion criteria, so the comparison concerns recorded disasters rather than all harmful weather. A cell-day recall of 1.79\,\% for heat means that 1.79\,\% of the rasterized heat-impact cell-days overlap the evaluated mask. It does not mean that only 1.79\,\% of heat disasters are captured; the event-level quantity is defined separately in Appendix~\ref{app:capture}.

\noindent\textbf{Score plateaus.}
The daily-hazard score is stored on a fine $0.0005$ grid (Sect.~\ref{app:processing}, corrected from an earlier draft's $k/99$ estimate, which was based only on the truncated 2023 file), but the sweep still plateaus near the top of the range: any day whose value exceeds the entire $\sim$930-value reference sample receives the same maximum empirical rank, so a substantial share of the most extreme days tie at or near 1 regardless of storage precision, and thresholds from about 0.99 to 1 select nearly the same top set. This is a property of the empirical-rank estimator's finite reference sample, not of how finely the result is stored. For drought, finite scores begin at approximately $1/3$, and non-drought scores are missing, explaining the identical selections at 0.1, 0.2, and 0.3. The drought score is capped at 1 for SPI values at or below $-3$. These different transformations prohibit interpreting a common numeric threshold as a common climatological percentile.

\noindent\textbf{Threshold selection and partial curve area.}
The reported $q^{\ast}$ is selected on the comparison data, so it describes that sample. Claims of generalization require independent evaluation. The partial area under the precision--recall curve integrates only over each sweep's attained recall range, which differs among hazards. It should not be used to rank hazards or compared directly with the full-curve no-skill baseline. Figure~\ref{fig:headline} retains this diagnostic for traceability. The rise of precision with threshold is an empirical result, whereas non-increasing recall follows from nested threshold masks when persistence and validity rules preserve nesting.

\section{Event-level capture rates}
\label{app:capture}

The concordance assessment of Sect.~\ref{sec:quality} is computed over grid
cell-days. A complementary event-level view asks, for each catalog record
separately, whether the tail label fired anywhere inside that record's own
footprint during its own recorded dates. Because EM-DAT frequently records only
a month or a year rather than a day, the date window is precision-aware --
$\pm2$ days for day-precise records, $\pm15$ for month-precise, and $\pm182$ for
the rare year-precise ones -- rather than a uniform $\pm2$ days, which would
systematically miss the true date of an imprecisely dated event. Capture rates
so defined are 55.3\,\% for heat ($n=219$), 76.7\,\% for cold
($n=288$), 81.8\,\% for wind ($n=2495$), 93.3\,\% for heavy
precipitation ($n=3966$) and 92.5\,\% for drought ($n=402$). Moving from a
uniform to a precision-aware window raised the heat and cold rates substantially
(from 46.6 and 63.2\,\%), demonstrating sensitivity to the matching window without identifying the cause of every unmatched event
(Sect.~\ref{sec:capture}).

Two stratifications qualify these numbers. First, capture rate rises with
footprint area for all five hazards, from 31\,\% to 86\,\% across
area quartiles for heat, and 60\,\% to 88\,\% for wind, simply
because a larger polygon evaluated over a multi-day window offers more
opportunities for some cell to cross the threshold somewhere, irrespective of
whether the exceedance coincides with the damage. This is the country-size
confound of Sect.~\ref{sec:assessment-caveats} reappearing as metric inflation
for large events. Second, stratifying by recorded severity gives only a mixed
signal, capture rises with severity for wind, heavy precipitation, and drought
but not for heat or cold.

An illustrative independent-cell calculation assigns chance capture probability $1-(1-p)^{AD}$ to a footprint containing $A$ cells over $D$ days, where $p$ is the tail rate at the evaluated threshold. The supplied figure uses representative areas and durations. This is not a calibrated null model for spatially coherent, persistent weather events. In particular, a constant conversion of approximately \SI{540}{\kilo\meter\squared} per cell is not valid across a latitude--longitude grid; the number of rasterized cells should be counted directly.

The supplied calculation reports observed-minus-null differences of 22, 51, and 61 percentage points for heat, wind, and precipitation in the smallest area quartile. These values describe this simplified calculation and do not establish statistical significance or causal agreement. Large footprints approach certain capture under the independent-cell model, illustrating the sensitivity of the metric to area and duration.

\begin{figure*}[tbp]
\centering
\includegraphics[width=\textwidth]{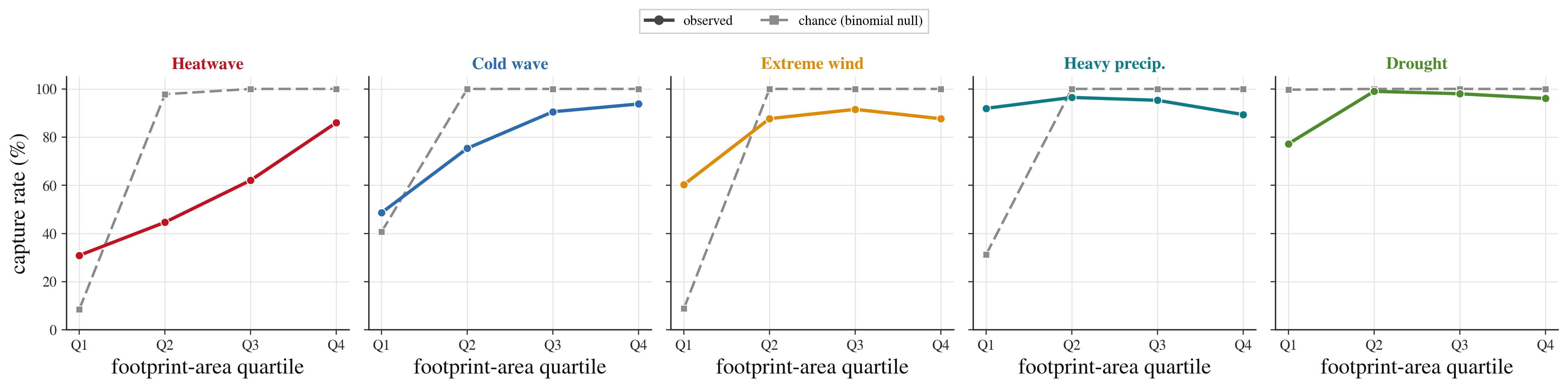}
\caption{\textbf{Event-level capture against a chance baseline, by footprint-area quartile.} Observed capture (solid, per hazard) and the binomial-null chance
capture (dashed) for a footprint of the quartile's representative area over a
representative $D$-day window, evaluated at each hazard's $q^{\ast}$. For
localized events (Q1) observed capture exceeds chance for heat, wind and heavy
precipitation, under this illustrative independence calculation; for large footprints (Q4) the chance rate saturates near 100\,\%, illustrating strong footprint-size sensitivity. Drought capture is at or below this illustrative baseline; interpretation requires a dependence-preserving null.}
\label{fig:capture-null}
\end{figure*}

\section{Per-hazard concordance detail}
\label{app:curves}

\begin{figure*}[tbp]
\centering
\includegraphics[width=\textwidth]{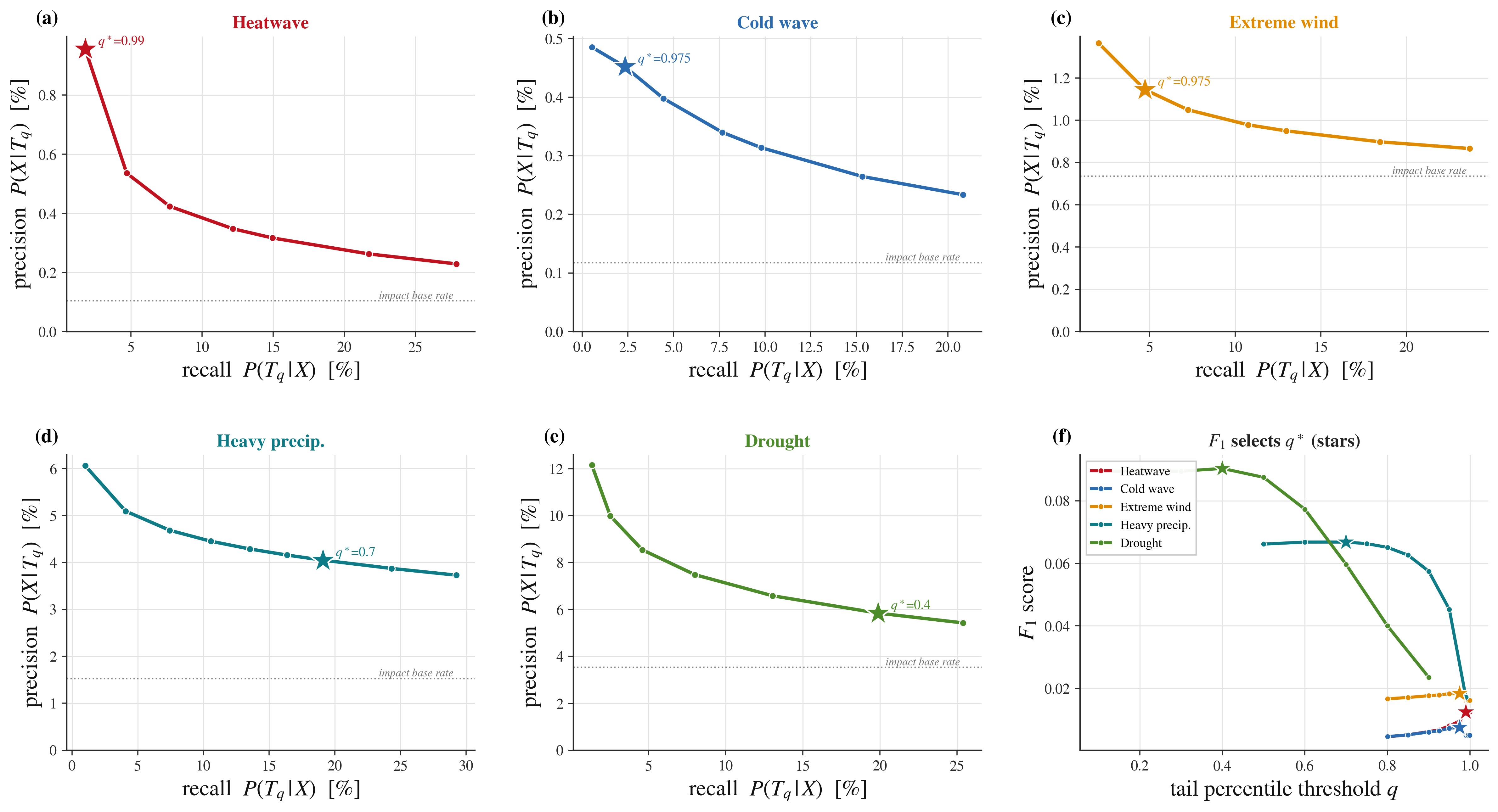}
\caption{\textbf{Per-hazard precision--recall curves.} (a--e) Precision
$P(X\mid T_q)$ against recall $P(T_q\mid X)$ as the tail threshold $q$ is swept,
one plot per hazard; the star marks the $F_1$-optimal $q^{\ast}$ and the dotted
line the impact base rate (the precision a random tail would achieve). (f) The
$F_1$ score as a function of $q$ for all five hazards, with stars at the maxima
that define $q^{\ast}$; these fall anywhere from $q^{\ast}=0.40$ (drought) to
$0.99$ (heat).}
\label{fig:pr-all}
\end{figure*}

\begin{figure*}[tbp]
\centering
\includegraphics[width=0.93\textwidth]{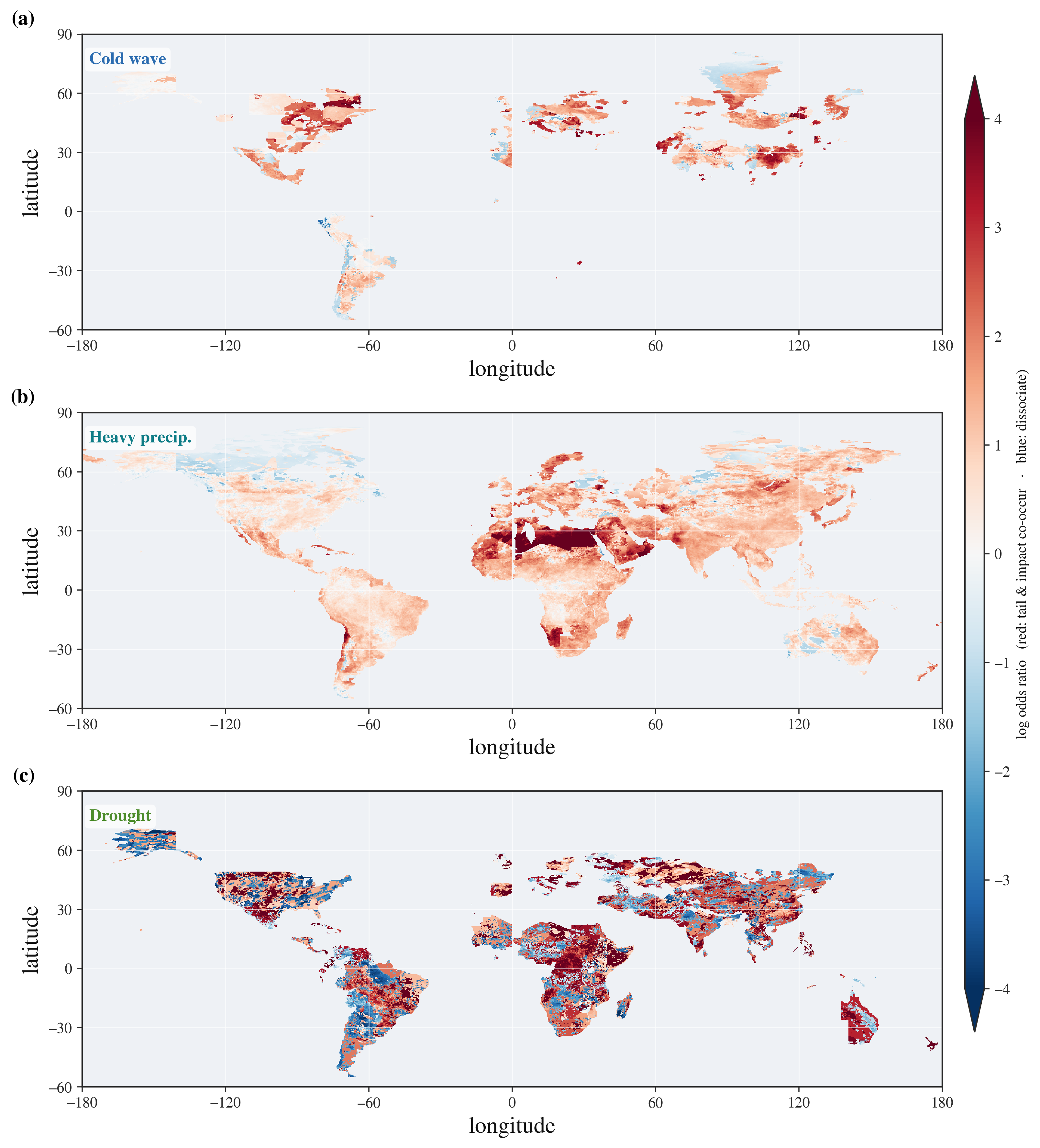}
\caption{\textbf{Per-cell log-odds ratio for the remaining three hazards} at
$q=0.90$, 1996--2023, as in Fig.~\ref{fig:map-heat}: (a) cold wave, (b) heavy
precipitation, (c) drought. Cells with fewer than five impact days are masked
and the diverging scale is clipped at $\pm4$. Heavy precipitation is estimable
over the widest area, its impact record being the densest.}
\label{fig:map-others}
\end{figure*}

The pooled concordance results of Sect.~\ref{sec:quality} are resolved here by
hazard. Figure~\ref{fig:pr-all} traces the precision--recall trade-off swept by
the tail threshold $q$ for each hazard, with the $F_1$-optimal $q^{\ast}$ marked.
Every curve sits well above its impact base rate (the precision a random tail
would achieve), indicating positive association in the reported sample,
while the $F_1$ maxima (plot~f) fall at very different thresholds, restating
that no single percentile is optimal. Figure~\ref{fig:map-others} completes the
spatial picture of Figs.~\ref{fig:map-heat} and \ref{fig:map-wind} with the
per-cell log-odds ratios for the three remaining hazards.

\section{Severity-tier composition through time}
\label{app:tiers}

\begin{figure*}[tbp]
\centering
\includegraphics[width=\textwidth]{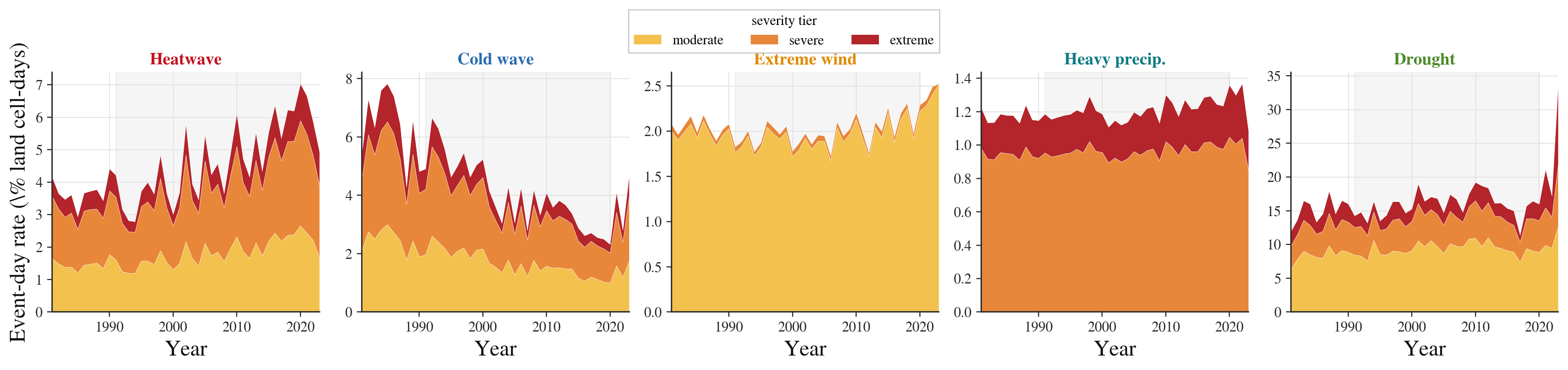}
\caption{\textbf{Severity-tier composition of the annual event-day rate,
1981--2022 plus a partial 2023 (see Fig.~\ref{fig:frequency}).} Each plot
stacks the moderate, severe, and extreme contributions
to the event-day rate of Fig.~\ref{fig:frequency} for one hazard, with the
1991--2020 reference period shaded. The vertical scale is independent per
hazard; the 2023 endpoint reflects a single partial month or ten days, not a
full year, most visibly for drought (Sect.~\ref{app:tiers}).}
\label{fig:tiers}
\end{figure*}

Figure~\ref{fig:frequency} reports the total event-day rate;
Fig.~\ref{fig:tiers} resolves it into the moderate, severe, and extreme tiers
year by year. The supplied frequency changes are tier-dependent: the heatwave
increase spans all three tiers but is proportionally largest in the extreme
tier, which roughly doubles between 1981 and 2020, and the plotted 2023 drought value has large severe and extreme contributions. Because only one January drought field is available in the inspected release, these contributions do not establish an annual drying signal. Heavy
precipitation carries no moderate tier by construction
(Sect.~\ref{sec:precip}), and extreme wind is almost entirely moderate because
its Beaufort escalation thresholds are seldom met.

\noappendix

\authorcontribution{Z.-S. Liu implemented the labeling pipeline, the concordance analysis, and the figures; M. Boy and R. Makkonen advised on atmospheric interpretation; all authors contributed to the manuscript.}

\competinginterests{The authors declare that they have no competing interests.}


\clearpage

\bibliographystyle{copernicus}
\bibliography{main}

\end{document}